\documentclass[12pt]{article}

\usepackage{graphicx}
\usepackage{float}

\def\gtwid{\mathrel{\raise.3ex\hbox{$>$\kern-.75em\lower1ex\hbox{$\sim$}}}}
\def\ltwid{\mathrel{\raise.3ex\hbox{$<$\kern-.75em\lower1ex\hbox{$\sim$}}}}
\def\square{\kern1pt\vbox{\hrule height 1.2pt\hbox{\vrule width 1.2pt\hskip 3pt
   \vbox{\vskip 6pt}\hskip 3pt\vrule width 0.6pt}\hrule height 0.6pt}\kern1pt}

\begin{document}

\begin{titlepage}

\begin{flushright}
UFIFT-QG-18-01
\end{flushright}

\vskip 2cm

\begin{center}
{\bf Structure Formation in Nonlocal MOND}
\end{center}

\vskip 1cm

\begin{center}
L. Tan$^{*}$ and R. P. Woodard$^{\dagger}$
\end{center}

\vskip .5cm

\begin{center}
\it{Department of Physics, University of Florida,\\
Gainesville, FL 32611, UNITED STATES}
\end{center}

\vspace{1cm}

\begin{center}
ABSTRACT
\end{center}
We consider structure formation in a nonlocal, metric-based realization
of Milgrom's MOdified Newtonian Dynamics (MOND). We derive the general 
equations for linearized scalar perturbations about the $\Lambda$CDM 
expansion history. These equations are considerably simplified for
sub-horizon modes, and it becomes obvious (in this model) that the MOND
enhancement is not sufficient to allow ordinary matter to drive 
structure formation. We discuss ways in which the model might be changed
to correct the problem.

\begin{flushleft}
PACS numbers: 04.50.Kd, 95.35.+d, 98.62.-g
\end{flushleft}

\begin{flushleft}
$^{*}$ e-mail: billy@ufl.edu \\
$^{\dagger}$ e-mail: woodard@phys.ufl.edu
\end{flushleft}

\end{titlepage}

\section{Introduction}

The failure of increasingly sensitive detectors to uncover any direct evidence 
for dark matter \cite{Aprile:2017iyp,Cui:2017nnn} has motivated a re-examination
of the dark matter paradigm. At the same time, powerful observational evidence
\cite{McGaugh:2016leg,Lelli:2017vgz} continues to accumulate in support of the 
predictions of Milgrom's MOdified Newtonian Dynamics (MOND) \cite{Milgrom:1983ca,
Milgrom:1983pn,Milgrom:1983zz} for galaxies. It seems that either MOND is correct 
or else the evolution of dark matter systems is driven towards a yet unrecognized
attractor solution which reproduces the predictions of MOND.

MOND can be viewed as the nonrelativistic, static limit of some modified 
gravity theory in which cosmic motions are explained without the need for dark 
matter. Reconstructing this theory has proven to be challenging. The only local, 
stable, metric-based extensions of general relativity are $f(R)$ models 
\cite{Woodard:2006nt}, and these cannot reproduce MOND. We must therefore either 
abandon locality or give up the metric as the exclusive carrier of the 
gravitational force. 

The first partially successful relativistic generalization of MOND was 
Bekenstein's TeVeS \cite{Bekenstein:2004ne}, which introduces scalar and vector 
fields to carry the extra gravitational force needed in the absence of dark
matter. The creation of TeVeS quickly belied some of the facile pronouncements 
that had hitherto been made about the impossibility of an alternative to dark 
matter. For example, TeVeS does a better job of explaining large scale structure 
than had been thought possible \cite{Skordis:2005xk,Skordis:2005eu,Dodelson:2006zt,
Bourliot:2006ig,Zlosnik:2007bu}. The model does have some potential issues with
stability \cite{Contaldi:2008iw}, and suffers from the need for fine tuning to
accommodate pulsar timing data \cite{Freire:2012mg}. More seriously, the model
fails to agree with detailed measurements of lensing, clustering and structure 
growth \cite{Reyes:2010tr}, and it generically produces baryon acoustic 
oscillations which are too large \cite{Dodelson:2011qv}. The ``gold-plated event''
needed to completely falsify TeVeS was recently provided by the nearly simultaneous
observation of gravitational wave and electromagnetic signals from a binary neutron
star merger about 40 Mpc away \cite{Boran:2017rdn}. At this distance TeVeS predicts
that gravitational waves should arrive hundreds of days before  electromagnetic 
radiation \cite{Desai:2008vj}.

Local relativistic generalizations of MOND which avoid the fatal problem of 
different arrival times employ other fields to change the gravitational field 
equations, but still use the metric to carry the gravitational force. A special
type of Einstein-Aether models falls into this category \cite{Zlosnik:2006zu,
Sanders:2011wa,Blanchet:2011wv}. A completely different class of models is
provided by Milgrom's bimetric formulation \cite{Milgrom:2009gv,Milgrom:2013iea}.

Fully metric-based generalizations of MOND cannot be local. Nonlocality is
problematic in fundamental theory but it may be that MOND emerges from the 
nonlocal effective field equations which describe the gravitational vacuum 
polarization induced by primordial inflation \cite{Woodard:2014wia}. This 
proposal is supported by the fact that loop corrections to the gravitational 
potentials during inflation show a small long range enhancement 
\cite{Wang:2015eaa,Park:2015kua,Frob:2016fcr}. A purely phenomenological model 
was devised which reproduces the Tully-Fisher relation and sufficient weak 
lensing \cite{Deffayet:2011sk,Deffayet:2014lba}. When this model's free 
function was tuned to reproduce most of the $\Lambda$CDM expansion history 
\cite{Kim:2016nnd} a serendipitous explanation appeared for the tension 
between determinations of the current expansion rate derived from data at 
high \cite{Ade:2015xua} and low \cite{Riess:2016jrr} redshift.

The purpose of this paper is to study cosmological perturbations. In section 
2 we review the model. Section 3 presents the equations for linearized scalar 
perturbations around the cosmological background. At the end of section 3 we 
make the sub-horizon approximation and study the growth of structure. Section 
4 discusses the implications for model-building.

\section{Past Work on the Model}

The purpose of this section is to review nonlocal MOND. We begin by giving
the field equations for a general geometry. These equations are then specialized
to static, spherically symmetric geometries, which reproduces the Tully-Fisher
relation and weak lensing. Finally, the general equations are specialized to
a homogeneous and isotropic background, which reproduces the $\Lambda$CDM
expansion history until a redshift of about $z_* \approx 0.088$.

\subsection{The Full Model}

Although the model is not local, its simplest formulation employs four 
auxiliary scalar fields to absorb the various nonlocal components after the 
technique of Nojiri and Odintsov \cite{Nojiri:2007uq}. The localized
Lagrangian depends upon the (spacelike) metric $g_{\mu\nu}$ and the auxiliary
scalars $\phi$, $\xi$, $\chi$ and $\psi$ \cite{Deffayet:2014lba},
\begin{eqnarray}
\lefteqn{\mathcal{L} = \frac{c^4}{16\pi G} \Biggl\{ R + \frac{a_0^2}{c^4} 
f_y\Bigl(\frac{g^{\mu\nu} \partial_{\mu} \phi \partial_{\nu} \phi}{
c^{-4} a_0^2} \Bigr) } \nonumber \\
& & \hspace{2cm} - \Bigl[ \partial_{\mu} \xi \partial_{\nu} \phi
g^{\mu\nu} \!+\! 2 \xi R_{\mu\nu} u^{\mu} u^{\nu} \Bigr] - \Bigl[
\partial_{\mu} \psi \partial_{\nu} \chi g^{\mu\nu} \!-\! \psi\Bigr]
\Biggr\} \sqrt{-g} \; , \qquad \label{Lagrangian}
\end{eqnarray}
where $a_0 \approx 1.2 \times 10^{-10}~{\rm m/s}^2$ is the characteristic
acceleration below which MOND phenomenology becomes apparent. The timelike 
4-velocity field $u^{\mu}$ is formed from the gradient of $\chi$,
\begin{equation}
u^{\mu} \equiv \frac{-g^{\mu\nu} \partial_{\nu} \chi[g]}{\sqrt{-
g^{\alpha\beta} \partial_{\alpha} \chi[g] \partial_{\beta} \chi[g]}} \; .
\label{udef}
\end{equation}
Regarding the four auxiliary scalars as independent fields would result in 
two ghosts \cite{Deser:2013uya,Woodard:2014iga}. The scalars must instead 
be regarded as nonlocal functionals of the metric which are obtained by 
solving their field equations using retarded boundary conditions,
\begin{eqnarray}
\phi[g] = \frac2{\square} R_{\alpha\beta} u^{\alpha} u^{\beta}  
&\!\! , \!\!& \chi[g] = -\frac1{\square} 1 \; , \label{scalars1} \\
\xi[g] = \frac2{\square} D^{\mu} \Biggl[ \partial_{\mu} \phi f'_y\Bigl(
\frac{ g^{\rho\sigma} \partial_{\rho} \phi \partial_{\sigma} \phi}{c^{-4} a_0^2}
\Bigr) \Biggr] &\!\! , \!\! & \psi[g] = \frac{4}{\square} D_{\mu} 
\Biggl[ \frac{ \xi (g^{\mu\rho} \!+\! u^{\mu} u^{\rho}) u^{\sigma} 
R_{\rho\sigma}}{\sqrt{-g^{\alpha\beta} \partial_{\alpha} \chi \partial_{\beta}
\chi}}\Biggr] . \quad \label{scalars2}
\end{eqnarray}
Here and henceforth $D_{\mu}$ denotes the metric-compatible covariant derivative.
The full gravitational field equations are,
\begin{eqnarray}
\lefteqn{ R_{\mu\nu} + \frac12 g_{\mu\nu} \Bigl[ -R -\frac{a_0^2}{c^4}
f_y + g^{\rho\sigma} \Bigl( \partial_{\rho} \xi \partial_{\sigma} \phi 
\!+\! \partial_{\rho} \psi \partial_{\sigma} \chi\Bigr) + 2 \xi 
u^{\rho} u^{\sigma} R_{\rho\sigma} - \psi\Bigr] } \nonumber \\
& & \hspace{-.3cm} + \partial_{\mu} \phi \partial_{\nu} \phi f_y' -
\partial_{(\mu} \xi \partial_{\nu)} \phi - \partial_{(\mu} \psi
\partial_{\nu)} \chi - 2 \xi \Bigl[ 2 u_{(\mu} u^{\alpha} R_{\nu ) \alpha}
\!+\! u_{\mu} u_{\nu} u^{\alpha} u^{\beta} R_{\alpha\beta} \Bigr]
\nonumber \\
& & \hspace{.3cm} - \Bigl[ \square (\xi u_{\mu} u_{\nu}) +
g_{\mu\nu} D_{\alpha} D_{\beta} (\xi u^{\alpha} u^{\beta}) -
2 D_{\alpha} D_{(\mu} (\xi u_{\nu)} u^{\alpha} ) \Bigr] 
= \frac{8\pi G}{c^4} \, T_{\mu\nu} \; . \qquad \label{graveqns}
\end{eqnarray}

The function $f_y(Z)$ which appears in equations (\ref{Lagrangian}), 
(\ref{scalars2}) and (\ref{graveqns}) is chosen to make the model agree
with phenomenology in three regimes. The first is for small, positive $Z$,
which corresponds to very weakly gravitationally bound systems whose
Newtonian gravitational acceleration is comparable to $a_0$ or smaller. 
In this regime the Tully-Fisher relation implies \cite{Deffayet:2011sk},
\begin{equation}
0 < Z \ll 1 \qquad \Longrightarrow \qquad f_y(Z) = \frac12 Z - 
\frac16 Z^{\frac32} + O(Z^2) \; . \label{TFrel}
\end{equation}
The second regime is for $Z \gtwid 1$, which corresponds to more strongly
gravitationally bound systems. Ensuring that general relativity applies 
in this regime requires $f_y(Z)$ to fall off rapidly. One function which
meets this requirement, and is also consistent with (\ref{TFrel}), is 
\cite{Deffayet:2014lba},
\begin{equation}
0 < Z < \infty \qquad \Longrightarrow \qquad f_y(Z) = \frac12 Z \,
\exp\Bigl[-\frac13 \sqrt{Z} \, \Bigr] \; . \label{posZ}
\end{equation}
The final regime is for negative $Z$, which corresponds to cosmology. In
this regime the function was numerically fit to reproduce the $\Lambda$CDM
expansion history, without dark matter, from very early times until a 
redshift of $z_* \approx 0.088$ \cite{Kim:2016nnd}. 

\subsection{Static, Spherically Symmetric Geometries}

The invariant element for static, spherically geometries can be expressed
in terms of two potentials usually termed $A(r)$ and $B(r)$,
\begin{equation}
g_{\mu\nu} dx^{\mu} x^{\nu} = -B(r) c^2 dt^2 + A(r) dr^2 + r^2 d\Omega^2 
\; . \label{staticgeom}
\end{equation}
Specializing the general relations (\ref{udef}) and (\ref{scalars1}-\ref{scalars2})
to this geometry implies that the timelike 4-velocity field is,
\begin{equation}
u^{\mu} = \frac{\delta^{\mu}_0}{\sqrt{B(r)}} \; . \label{staticu}
\end{equation}
The action of the scalar d'Alembertian on a function $F(r)$ is,
\begin{equation}
\square F(r) = \frac1{r^2 \sqrt{AB}} \frac{d}{dr} \Biggl[ r^2
\sqrt{\frac{B(r)}{A(r)}} F'(r) \Biggr] \; . \label{staticsquare}
\end{equation}
In the geometry (\ref{staticgeom}) the $00$ component and trace of the Ricci 
tensor are,
\begin{eqnarray}
R_{00} & = & \frac{B''}{2A} - \frac{B'}{4 A} \Bigl( \frac{A'}{A}
\!+\! \frac{B'}{B}\Bigr) + \frac{B'}{r A} \; , \label{staticR00} \\
R & = & -\frac{B''}{AB} + \frac{B'}{2AB} \Bigl( \frac{A'}{A} \!+\!
\frac{B'}{B} \Bigr) + \frac2{r A} \Bigl( \frac{A'}{A} \!-\! \frac{B'}{B}
\Bigr) + \frac{2 (A \!-\! 1)}{r^2 A} \; . \qquad \label{staticR}
\end{eqnarray}
Substituting expressions (\ref{staticu}-\ref{staticR00}) into the general
relations (\ref{scalars1}-\ref{scalars2}) gives $\psi = 0$ and simple 
relations for the derivatives of $\phi(r)$ and $\xi(r)$,
\begin{equation}
\phi'(r) = \frac{B'(r)}{B(r)} \qquad , \qquad \xi'(r) = 2 \phi'(r) \!\times\!
f_y'\Biggl( \frac{c^4 {B'}^2(r)}{a_0^2 A(r) B^2(r)} \Biggr) \; . 
\label{staticphixi}
\end{equation}
From (\ref{staticphixi}) we see that the argument of $f_y(Z)$ for this
geometry is,
\begin{equation}
Z = \frac{c^4 {B'}^2(r)}{a_0^2 A(r) B^2(r)} \; . \label{staticZ}
\end{equation}

We specialize the perfect fluid stress tensor,
\begin{equation}
T_{\mu\nu} = (\rho + p) u_{\mu} u_{\nu} + p g_{\mu\nu} \; ,
\end{equation}
to this geometry with energy density $\rho(r)$ and pressure $p(r)$. Setting
$\mu = \nu = 0$ in (\ref{graveqns}) gives,
\begin{equation}
\frac{r A'}{A^2} + 1 - \frac1{A} + \frac{a_0^2}{2 c^4} \, r^2 f_y - 
\frac{r^2 {B'}^2}{A B^2} \, f_y' - \frac{2}{\sqrt{A}} \partial_r \Bigl[
\frac{r^2 B'}{\sqrt{A} B} \, f_y' \Bigr] = \frac{8\pi G}{c^4} \, r^2 \rho \; . 
\label{static00}
\end{equation}
The $r r$ component of (\ref{graveqns}) implies,
\begin{equation}
\frac{r B'}{A B} - 1 + \frac1{A} - \frac{a_0^2}{2 c^4} \, r^2 f_y + 
\frac{r^2 {B'}^2}{A B^2} \, f_y' = \frac{8 \pi G}{c^4} \, r^2 p \; . 
\label{staticrr}
\end{equation} 
The other two nontrivial equations in this geometry are implied by stress-energy 
conservation.

If we assume the form (\ref{posZ}) for $f_y(Z)$ equations (\ref{static00}-\ref{staticrr})
can be reduced,
\begin{eqnarray}
\frac{8\pi G}{c^4} \, r^2 \rho & = & \Bigl[ r \Bigl(1 - \frac1{A}\Bigr)\Bigr]' -
\frac{r^2 {B'}^2}{4 A B^2} \Bigl[1 - \frac{c^2 B'}{3 a_0 \sqrt{A} B} \Bigr] 
\exp\Bigl[ -\frac{c^2 B'}{3 a_0 \sqrt{A} B} \Bigr] \nonumber \\
& & \hspace{1cm} - \frac{1}{\sqrt{A}} \partial_r \Biggl\{ \frac{r^2 B'}{\sqrt{A} B} 
\Bigl[1 - \frac{c^2 B'}{6 a_0 \sqrt{A} B}\Bigr] \exp\Bigl[- \frac{c^2 B'}{3 a_0 \sqrt{A} B}
\Bigr] \Biggr\} \; , \qquad \\
\frac{8 \pi G}{c^4} \, r^2 p & = & \frac{r B'}{A B} - 1 + \frac1{A} +
\frac{r^2 {B'}^2}{4 A B^2} \Bigl[1 - \frac{c^2 B'}{3 a_0 \sqrt{A} B} \Bigr] 
\exp\Bigl[ -\frac{c^2 B'}{3 a_0 \sqrt{A} B} \Bigr] \; . \label{staticrrred} 
\end{eqnarray}
To understand how the Tully-Fisher relation and weak lensing emerge, set the pressure
to zero and expand in the deviations,
\begin{equation}
a(r) \equiv A(r) - 1 \qquad , \qquad b(r) \equiv B(r) - 1 \; .
\end{equation}
Note also that the length $\frac{c^2}{a_0} \simeq 7.5 \times 10^{26}~{\rm m}$ is 
greater than the current Hubble radius, so quadratic terms which lack this length
can be discarded \cite{Soussa:2003vv}. Hence the equations become,
\begin{eqnarray}
\frac{8\pi G}{c^4} \, r^2 \rho & \simeq & \Bigl[r a - r^2 b' + 
\frac{c^2 r^2 {b'}^2}{2 a_0} \Bigr]' \; , \label{staticweak00} \\
0 & \simeq & r b' - a \; . \label{staticweakrr}
\end{eqnarray} 
Now use (\ref{staticweakrr}) --- which gives lensing --- to simplify 
(\ref{staticweak00}) and integrate, recognizing the mass $M(r)$ enclosed within
radius $r$,
\begin{equation}
\frac{c^2 r^2 {b'}^2}{2 a_0} \simeq \frac{2 G M(r)}{c^2} \quad \Longrightarrow 
\quad v_{\infty}^2 \equiv \frac12 c^2 \lim_{r \rightarrow \infty} r b'(r) =
\sqrt{a_0 G M(\infty)} \; . \label{TullyFisher}
\end{equation}
 
\subsection{Homogeneous and Isotropic Geometries}

The invariant element for a spatially flat, homogeneous and isotropic geometry
is,
\begin{equation}
ds^2 -c^2 dt^2 + a^2(t) d\vec{x} \!\cdot\! d\vec{x} \quad \Longrightarrow \quad
H(t) \equiv \frac{\dot{a}}{a} \; . \label{FLRW}
\end{equation}
With initial time $t_i$, auxiliary scalars for this geometry are 
\cite{Deffayet:2014lba},
\begin{eqnarray}
\phi_0(t) & \!\!\!=\!\!\! & 6 \!\! \int_{t_i}^t \!\! \frac{dt'}{a^3(t')} \!\!
\int_{t_i}^{t'} \!\! dt'' a^3(t'') \Bigl[H^2(t'') \!+\! \dot{H}(t'')\Bigr]
\; \Longrightarrow \; Z_0(t) = -\frac{c^2 \dot{\phi}^2(t)}{a_0^2} \; , \qquad 
\label{FLRWphi} \\
\chi_0(t) & \!\!\!=\!\!\! & \!\! \int_{t_i}^t \!\! \frac{dt'}{a^3(t')} \!\!
\int_{t_i}^{t'} \!\! dt'' a^3(t'') \qquad \Longrightarrow \qquad
u_0^{\mu}(t) = \delta^{\mu}_0 \; , \qquad \label{FLRWchi} \\
\xi_0(t) & \!\!\!=\!\!\! & 2 \!\! \int_{t_i}^t \!\! dt' \dot{\phi}_0(t') 
f_y'\Bigl(Z_0(t') \Bigr) \qquad , \qquad \psi_0(t) = 0 \; . \label{FLRWxipsi}
\end{eqnarray}
We have bestowed a subscript 0 on these quantities to indicate that they 
represent the background solutions around which perturbations will be 
developed in the next section. The background gravitational field equations 
are \cite{Deffayet:2014lba},
\begin{eqnarray}
3 H^2 \!+\! \frac{a_0^2}{2 c^2} f_y(Z_0) \!+\! 3 H \dot{\xi}_0 \!+\! 6 H^2 \xi_0 
& \!\!\!=\!\!\! & \frac{8\pi G}{c^2} \, \rho_0 \; , \label{FLRW00} \\
-2\dot{H} \!-\! 3H^2 \!-\! \frac{a_0^2}{2 c^2} \, f_y(Z_0) \!-\! \ddot{\xi}_0 
\!-\! \Bigl(\frac{\dot{\phi}_0}2 \!+\! 4 H\Bigr) \dot{\xi}_0 \!-\! 
\Bigl(4\dot{H} \!+\! 6 H^2\Bigr) \xi_0  & \!\!\!=\!\!\! &  \frac{8\pi G}{c^2} 
\, p_0 \; . \qquad \label{FLRWij}
\end{eqnarray}
Here $\rho_0(t)$ and $p_0(t)$ are the background energy density and pressure 
without dark matter, which we parameterize using the cosmological redshift $z$,
\begin{eqnarray}
\rho_0(t) & \equiv & \frac{3 c^2 H_0^2}{8\pi G} \Bigl[ \Omega_r (1 + z)^4 + 
\Omega_b (1 + z)^3 + \Omega_{\Lambda}\Bigr] \; , \label{FLRWrho} \\
p_0(t) & \equiv & \frac{3 c^2 H_0^2}{8 \pi G} \Bigl[ \frac13 \Omega_r (1 + z)^4 -
\Omega_{\Lambda}\Bigr] \; , \label{FLRWp} \\
1 + z & \equiv & \frac{a(t_0)}{a(t)} \; . \label{redshift}
\end{eqnarray}
Whereas $\Omega_r$ and $\Omega_{\Lambda}$ are the $\Lambda$CDM values for the
fraction of critical density in radiation and vacuum energy, respectively, 
$\Omega_b$ is only the fraction of critical density in baryons. The fraction in
nonrelativistic matter is $\Omega_m = \Omega_c + \Omega_b$, where $\Omega_c
\approx 5.3 \times \Omega_b$ is the fraction in dark matter.

For this cosmological regime the argument $Z$ of the function $f_y(Z)$ is negative.
How the function $f_y(Z)$ depends on negative $Z$ is not fixed by MOND phenomenology. 
In a previous work \cite{Kim:2016nnd} we showed how the function $f_y(Z)$ could be
defined for $Z < 0$ to exactly reproduce the $\Lambda$CDM expansion, from very early 
times all the way up until a redshift of $z_* \approx 0.088$. The $\Lambda$CDM Hubble 
parameter rescaled by its current value $H_0$ is,
\begin{equation}
\widetilde{H}(z) = \sqrt{ \Omega_r (1 \!+\! z)^4 + \Omega_m (1 \!+\! z)^3 + 
\Omega_{\Lambda}} \; . \label{FLRWHubble}
\end{equation}
The process of constructing $f_y(Z)$ to support $\widetilde{H}(z)$ without dark 
matter proceeds in two steps. We begin the first step by introducing the variables,
\begin{equation}
f(z) \equiv -\frac{a_0^2 f_y(Z_0)}{36 c^2 H_0^2 \Omega_c} \; , \;
s(z) \equiv \frac{a_0 \sqrt{-Z_0}}{6 c H_0} \; , \; 
g(z) \equiv \int_{z}^{\infty} \!\! \frac{f'(\zeta) d\zeta}{(1 \!+\! \zeta) s'(\zeta) 
\widetilde{H}(\zeta)} \; . \label{fsg} 
\end{equation}
The first step consists of solving the following integral-differential equation for
$f(z)$,
\begin{equation}
\frac12 f(z) + \frac{\widetilde{H}(z) f'(z)}{2 s'(z)} + \widetilde{H}^2(z) \,
g(z) = \frac1{12} (1 \!+\! z)^3 \; . \label{feqn}
\end{equation}
In the second step we invert the relation for $s(z)$,
\begin{equation}
s(z) = (1 \!+\! z)^3 \! \int_{z}^{\infty} \!\! d\zeta \Biggl[\frac{(1 \!+\! \zeta) 
\widetilde{H}'(\zeta) - \widetilde{H}(\zeta)}{(1 \!+\! \zeta)^4} \Biggr] \; .
\label{sint}
\end{equation}
to determine the redshift $z$ as a function of $Z_0$. Then regarding $f(z)$ as a 
function of $Z_0$ gives $f_y(Z)$ through relation (\ref{fsg}). Although the function 
$s(z)$ is positive for large $z$, the presence of a cosmological constant causes it 
to vanish for $z = z_*$ \cite{Kim:2016nnd}, so we can only recover the $\Lambda$CDM 
expansion history for $z_* < z < \infty$.

Relation (\ref{feqn}) might seem to have two homogeneous solutions but it in fact 
has only one. To see this note first from expressions (\ref{FLRWHubble}) and 
(\ref{sint}) that the large $z$ forms of $\widetilde{H}(z)$ and $s(z)$ are,
\begin{equation}
\widetilde{H}(z) \longrightarrow \sqrt{\Omega_r} \, (1 \!+\! z)^2 \qquad , \qquad
s(z) \longrightarrow \sqrt{\Omega_r} \, (1 \!+\! z)^2 \; . \label{asympHs}
\end{equation}
Now consider a homogeneous solution of (\ref{feqn}) with an asymptotic large $z$ 
behavior of $(1 + z)^x$. If we assume $x < 4$ then the integral for $g(z)$ exists 
and we find an equation for the exponent $x$,
\begin{equation} 
f(z) \longrightarrow (1 \!+\! z)^x \;\; (x < 4) \;\; \Longrightarrow \frac12
+ \frac14 x + \frac{\frac12 x}{4 \!-\! x} = 0 \Longrightarrow x_{\pm} = 
2 (1 \pm \sqrt{3}) \; . \label{homogeneous}
\end{equation}
The solution $x_-$ is indeed less than 4, so it corresponds to a genuine 
homogeneous solution. However, $x_+ > 4$, which means the derivation of this 
solution is not self-consistent. Hence there is only a single homogeneous solution,
and equation (\ref{feqn}) requires a single condition. Because the positive $Z$ 
function (\ref{TFrel}) vanishes for $Z = 0$ it is natural to take this to be 
$f(z_*) = 0$.

\begin{figure}[H]
\includegraphics[width=6.0cm,height=4.0cm]{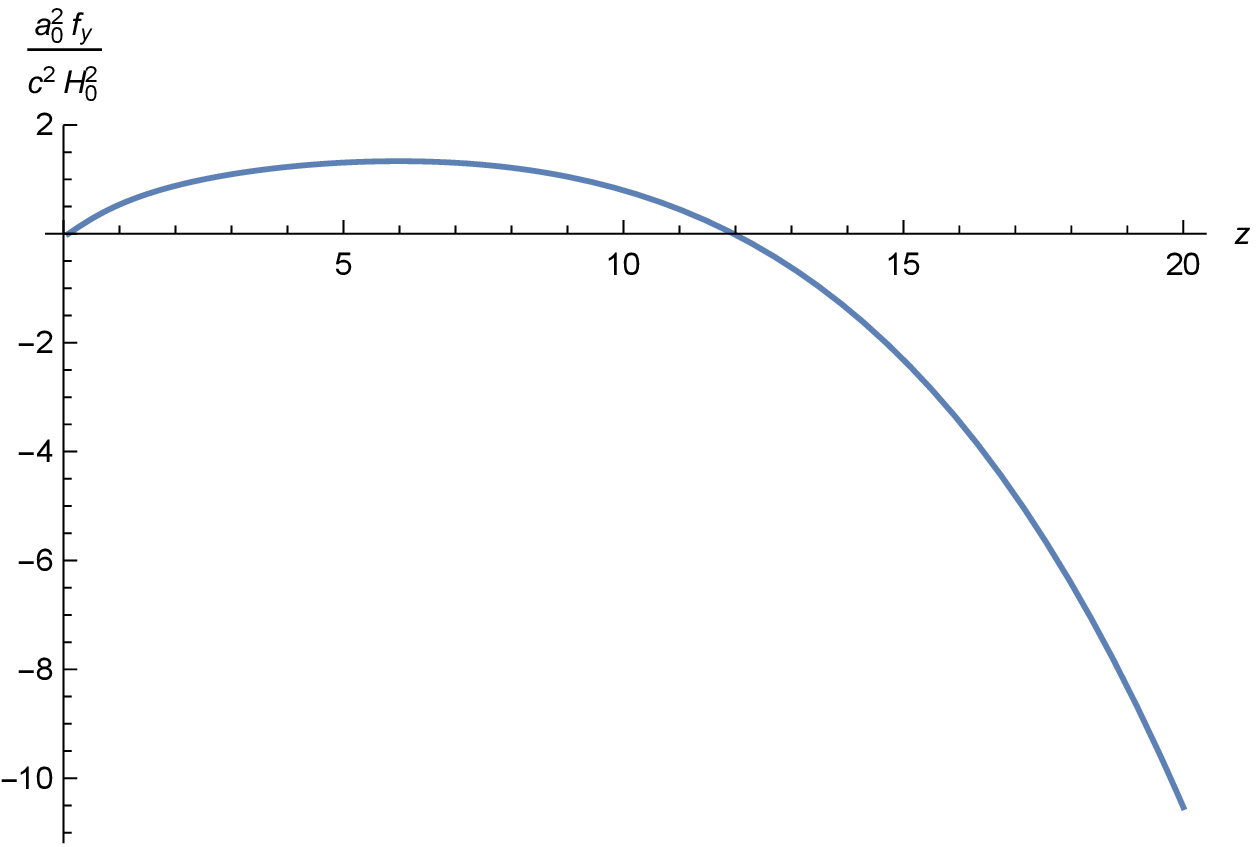}
\hspace{1cm}
\includegraphics[width=6.0cm,height=4.0cm]{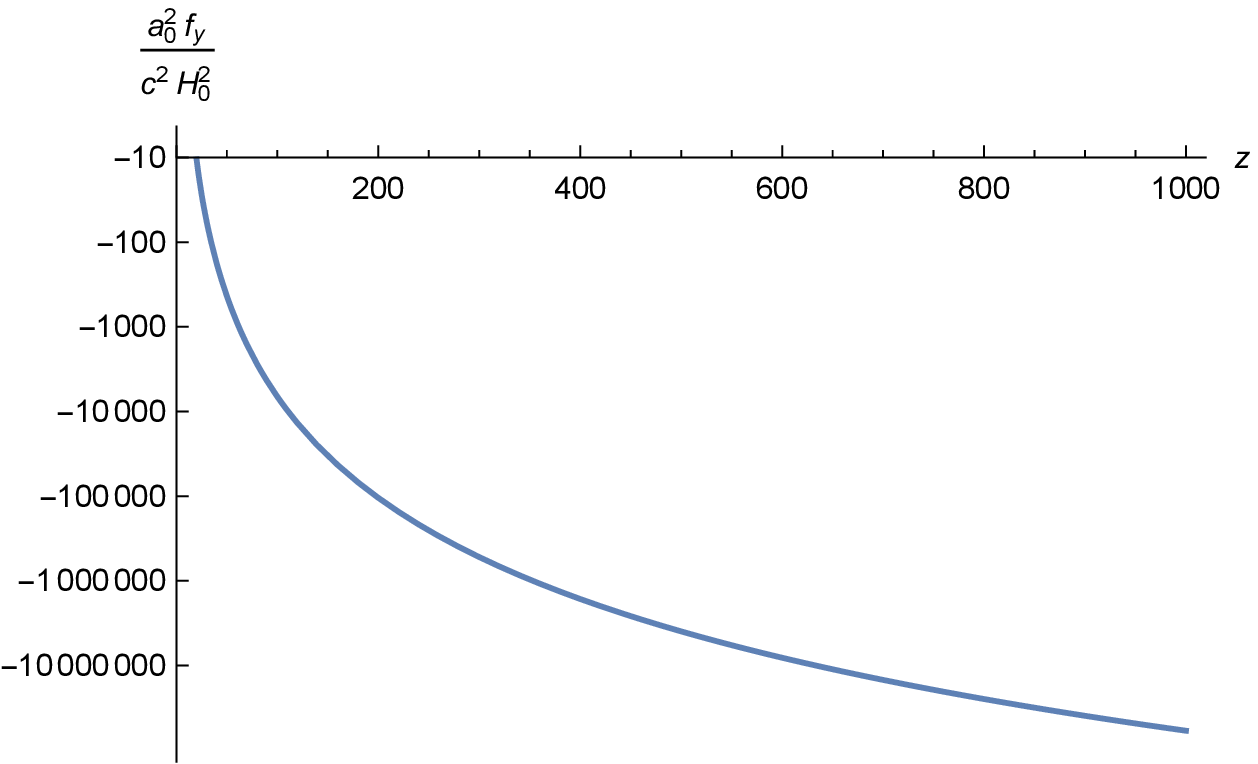}
\caption{Both graphs show the quantity $(\frac{a_0}{c H_0})^2 \times
f_y(Z_0)$ as a function of the redshift $z$. The left hand graph shows
that $f_y(Z_0)$ vanishes at $z = z_* \approx 0.088$ and is positive 
for $z_* < z \ltwid 14$. The right hand graph shows that $f_y(Z_0)$ is
negative, and monotonically decreasing, for all $z \gtwid 12$.}
\label{fy}
\end{figure}

A number of background quantities appear in the perturbation equations,
\begin{eqnarray}
\frac{a_0^2}{c^2 H_0^2} \, f_y(Z_0) = -36 \Omega_c \times f(z) & , & 
f'_y(Z_0) = \frac12 \Omega_c \times \frac{f'(z)}{s(z) s'(z)} \; , 
\label{fyfy'} \\
\xi_0(t) = -6\Omega_c \times g(z) & , & \frac{\dot{\xi}_0(t)}{H_0} = -6 \Omega_c
\times \frac{f'(z)}{s'(z)} \; , \label{xixidot} \\
\frac{\dot{\phi}_0(t)}{H_0} = -6 \times s(z) & , & H_0 \dot{\chi}_0(t) =
(1 \!+\! z)^3 \! \int_{z}^{\infty} \!\! \frac{d\zeta}{(1 \!+\! \zeta)^4 
\widetilde{H}(\zeta)} \; . \qquad \label{phidotchidot}
\end{eqnarray}
It is important to know the signs, magnitudes and rough $z$ dependences of
these quantities. Their behaviors for large $z$ follow from relation 
(\ref{asympHs}) for $s(z)$ and the implied relations $f(z) \longrightarrow 
\frac1{33} (1 \!+\! z)^3$ and $g(z) \longrightarrow \frac1{22 \Omega_r} 
\frac1{1 \!+\! z}$,
\begin{eqnarray}
\frac{a_0^2}{c^2 H_0^2} \, f_y(Z_0) \longrightarrow -\frac{12}{11} \Omega_c 
(1 \!+\! z)^3 & , & f'_y(Z_0) \longrightarrow \frac{\Omega_c}{44 \Omega_r} 
\frac{1}{1 \!+\! z} \; , \label{asympfyfy'} \\
\xi_0(t) \longrightarrow -\frac{3 \Omega_c}{11 \Omega_r} \frac1{1 \!+\! z} 
& , & \frac{\dot{\xi}_0(t)}{H_0} \longrightarrow -\frac{3\Omega_c}{11\sqrt{\Omega_r}}
\, (1 \!+\! z) \; , \label{asympxixidot} \\
\frac{\dot{\phi}_0(t)}{H_0} \longrightarrow -6 \sqrt{\Omega_r} \, (1 \!+\! z)^2 
& , & H_0 \dot{\chi}_0(t) \longrightarrow \frac1{5 \sqrt{\Omega_r}} 
\frac1{(1 \!+\! z)^2} \; . \qquad \label{asympphidotchidot}
\end{eqnarray}

\begin{figure}[H]
\includegraphics[width=6.0cm,height=5cm]{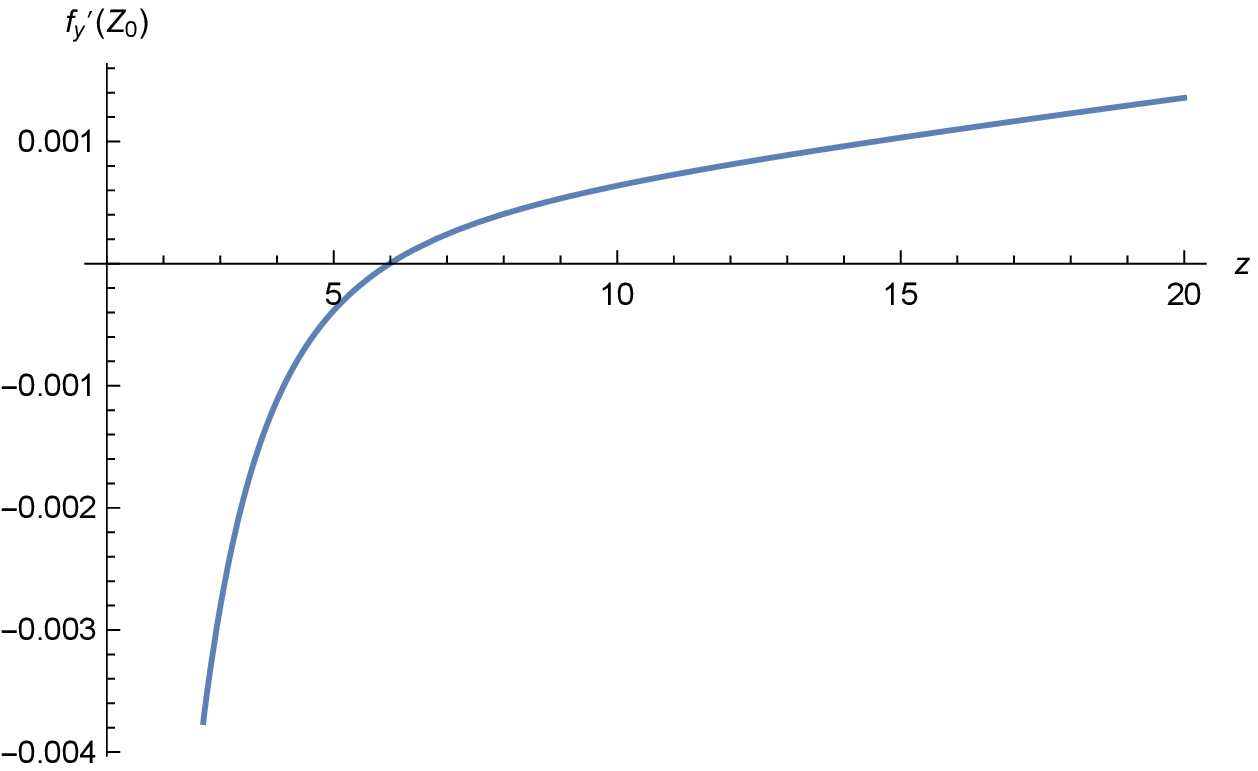}
\hspace{1cm}
\includegraphics[width=6.0cm,height=5cm]{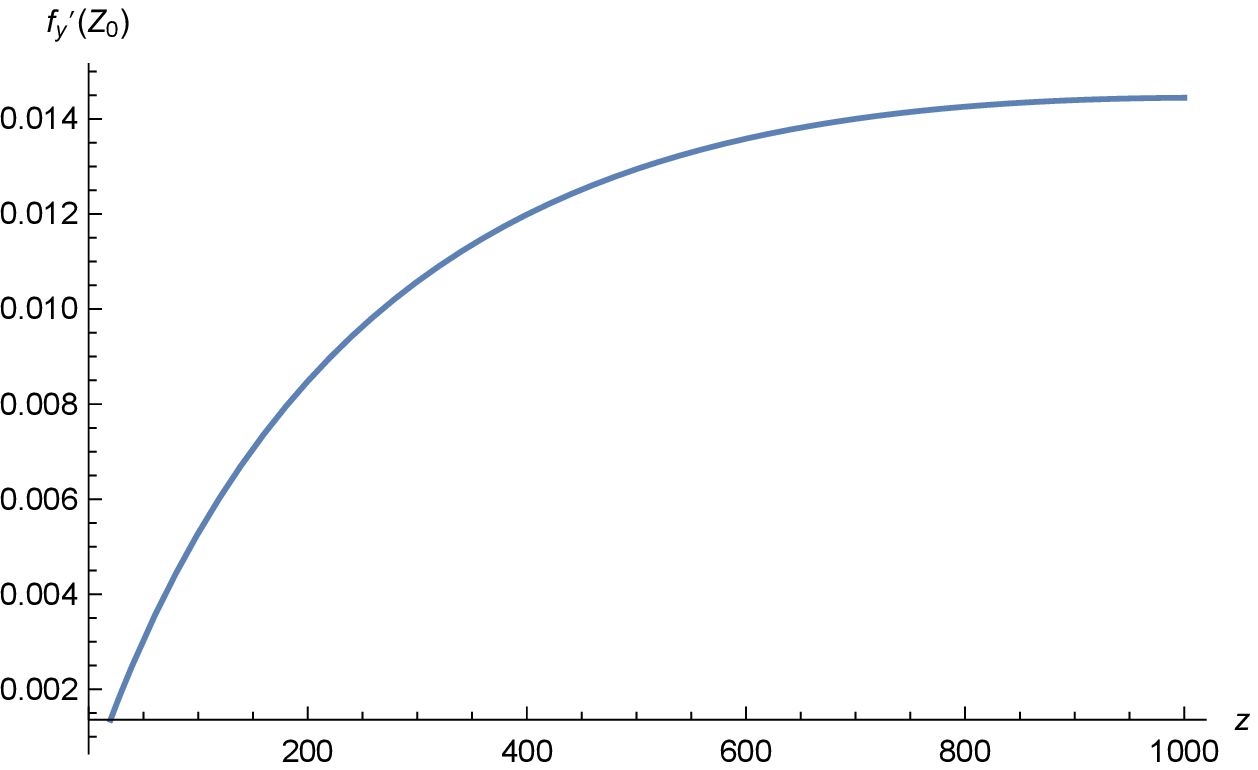}
\caption{Both graphs depict the quantity $f'_y(Z_0)$ as a function of the redshift
$z$. The left hand graph shows that $f_y'(Z_0)$ is negative for $z \ltwid 6$, and
approaches $-\infty$ at $z = z_* \approx 0.088$. The right hand graph demonstrates
that $f'_y(Z_0)$ never becomes larger than about 0.015.}
\label{fyprime}
\end{figure}

\begin{figure}[H]
\includegraphics[width=6.0cm,height=4.5cm]{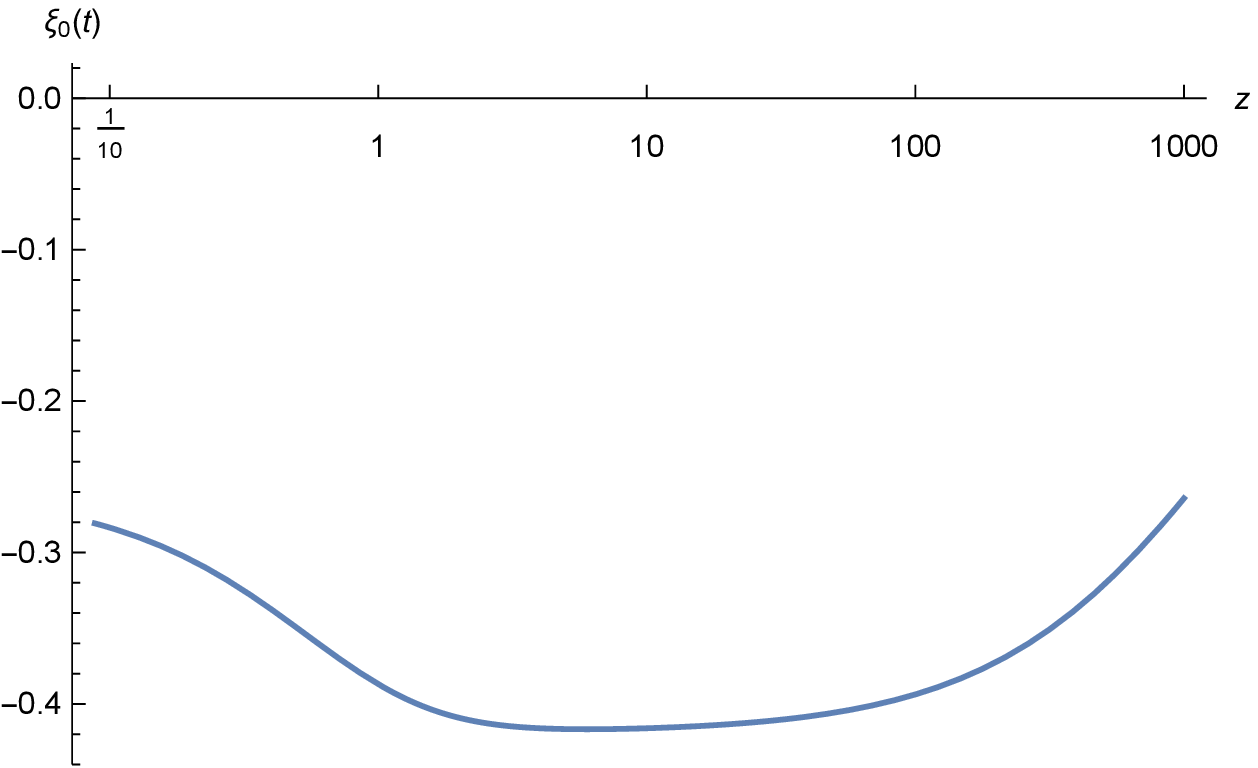}
\hspace{1cm}
\includegraphics[width=6.0cm,height=4.5cm]{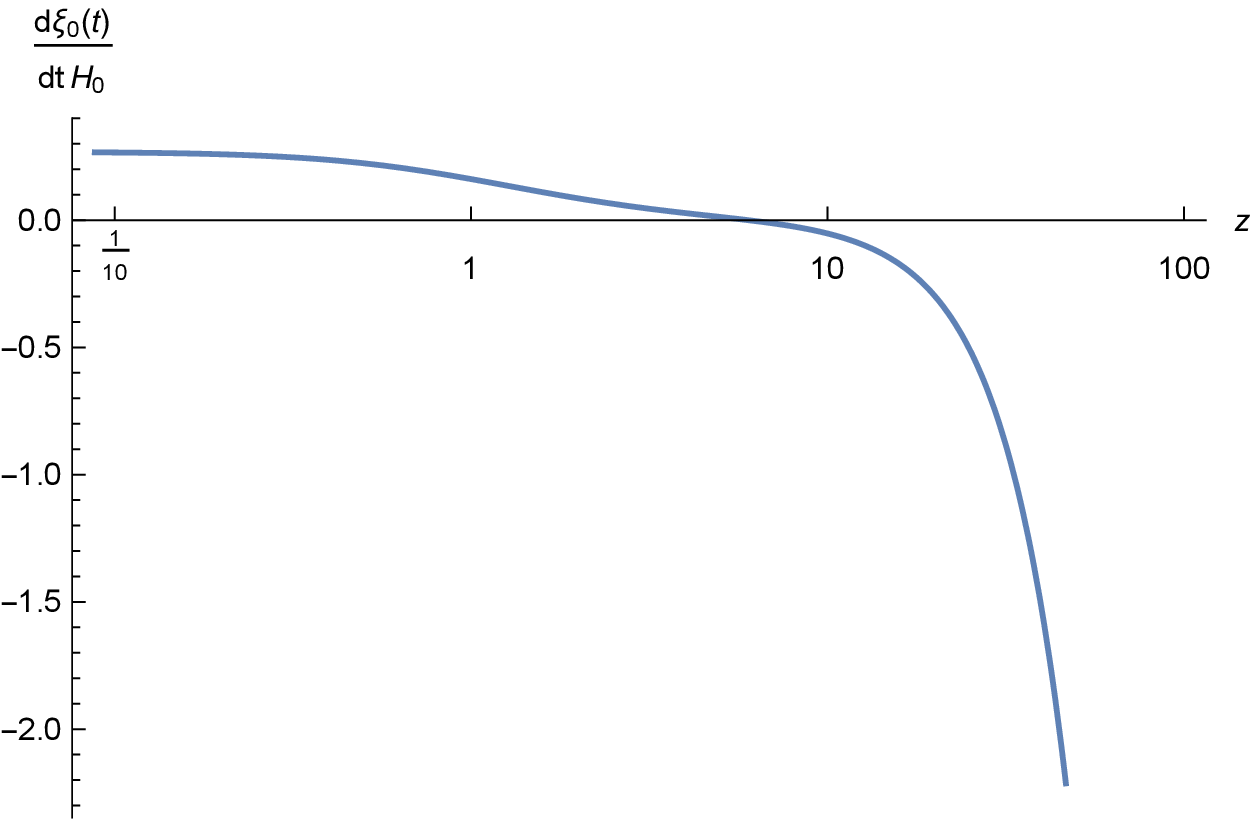}
\caption{These graphs show $\xi_0(t)$ (left) and $\dot{\xi}_0(t)/H_0$ (right) as
functions of the redshift $z$. Although $\xi_0(t)$ is negative definite, the sign
of $\dot{\xi}_0(t)/H_0$ is positive for $z_* < z \ltwid 6$ and negative for 
$6 \ltwid z < \infty$.}
\label{xixidotfigs}
\end{figure}

\begin{figure}[H]
\includegraphics[width=6.0cm,height=4.0cm]{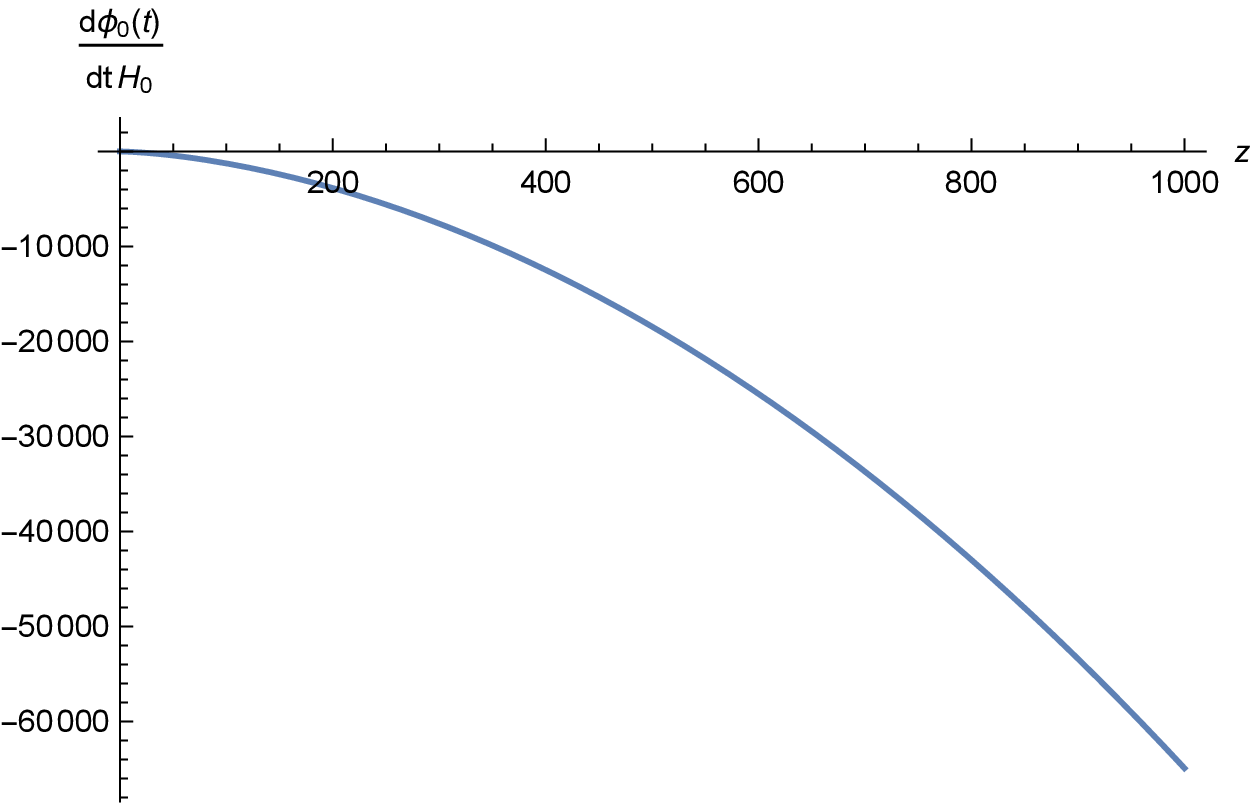}
\hspace{1cm}
\includegraphics[width=6.0cm,height=4.0cm]{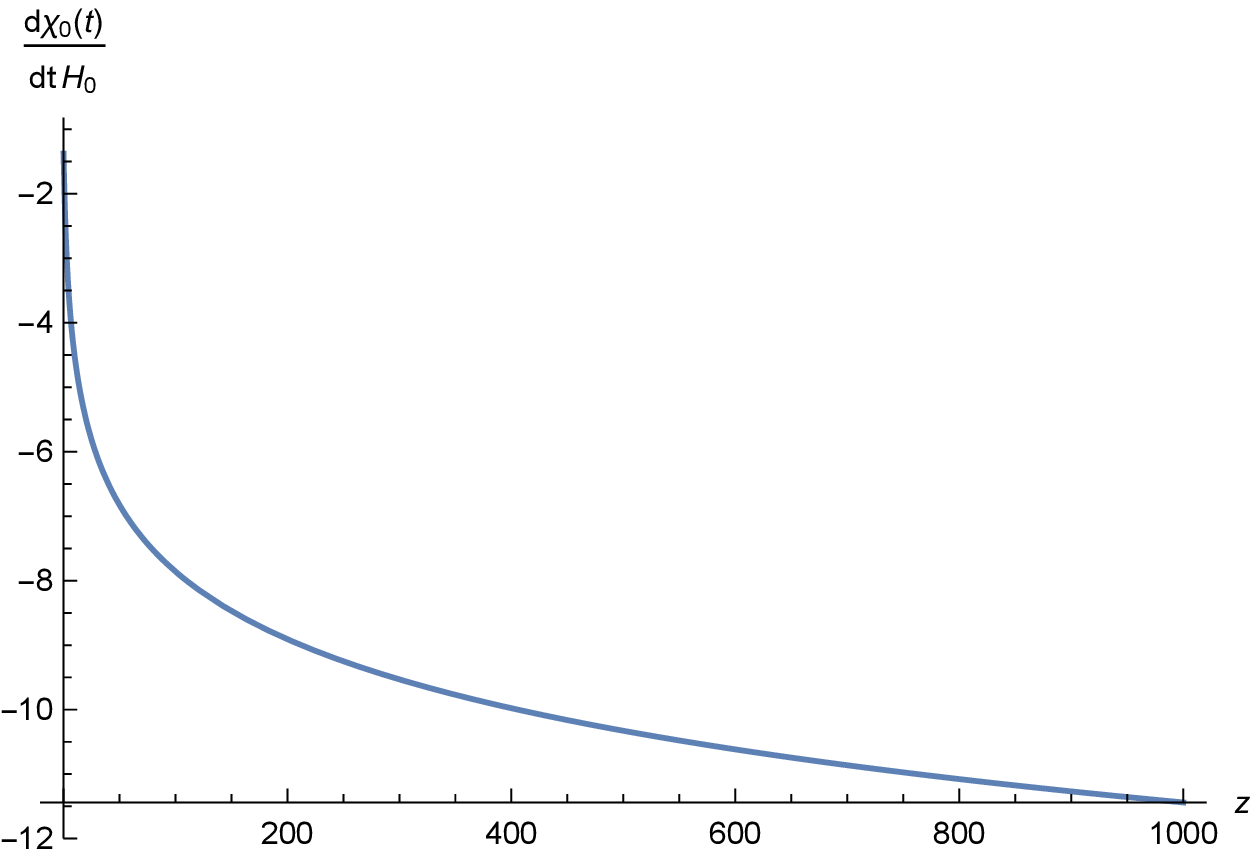}
\caption{These graphs show $\dot{\phi}_0(t)/H_0$ (left) and $\dot{\chi}_0(t)/H_0$ 
(right) as functions of the redshift $z$. The function $\dot{\phi}_0$ vanishes at
$z = z_*$; it is negative and monotonically decreasing for $z_* < z < 1000$. The
function $\dot{\chi}_0$ is positive and monotonically decreasing for $z_* \ge z \ge 
1000$.}
\label{phidotchidotfigs}
\end{figure}

Although the large $z$ results (\ref{asympfyfy'}-\ref{asympphidotchidot}) are 
valid, there is no alternative to numerically evaluating the six quantities 
(\ref{fyfy'}-\ref{phidotchidot}) for the redshifts of relevance to structure
formation. Fig.~\ref{fy} shows the function $f_y(Z_0)$ --- rescaled by 
$(\frac{a_0}{c H_0})^2$ --- as a function of the redshift $z$. Fig.~\ref{fyprime}
depicts $f_y'(Z_0)$ versus $z$, demonstrating both that the function approaches
$-\infty$ at $z = z_*$, and that it never becomes more positive than about 0.015 
in the range of interest for structure formation. That turns out to be the crucial
point in dooming this model. Figures~\ref{xixidotfigs} and \ref{phidotchidotfigs}
show the other relevant auxiliary scalars.

\section{Linearized Scalar Perturbations}

The purpose of this section is to give the equations for linearized,
scalar perturbations about the cosmological background of the previous
section. We begin by reviewing the well known geometrical relations, 
then move to giving the equations for the auxiliary scalars and for 
the geometry. Finally, the various relations are specialized to the
sub-horizon regime in which the comoving wavelength is much smaller 
than the Hubble length. To simplify the many tedious manipulations we 
work in units for which $c = 1$.

\subsection{Perturbed Geometry}

In Newtonian gauge the geometry of linearized scalar perturbations is,
\begin{equation}
g_{\mu\nu}(t,\vec{x}) dx^{\mu} dx^{\nu} = - \Bigl[1 \!+\! 2 
\widetilde{\Psi}(t,\vec{x}) \Bigr] dt^2 + a^2(t) \Bigl[1 \!+\! 2 
\widetilde{\Phi}(t,\vec{x}) \Bigr] d\vec{x} \!\cdot\! d\vec{x} 
\; . \label{pertmetric}
\end{equation}
Of course the perturbation fields can be decomposed into spatial plane waves,
\begin{equation}
\widetilde{\Psi}(t,\vec{x}) \equiv \int \!\! \frac{d^3k}{(2\pi)^3} \, 
e^{i \vec{k} \cdot \vec{x}} \, \Psi(t,\vec{k}) \qquad , \qquad 
\widetilde{\Phi}(t,\vec{x}) \equiv \int \!\! \frac{d^3k}{(2\pi)^3} \, 
e^{i \vec{k} \cdot \vec{x}} \, \Phi(t,\vec{k}) \; . \label{Fourier}
\end{equation}
Because we are only interested in the {\it linearized} evolution equations
each plane wave mode can be considered separately, and we will only report
results for these Fourier components. For example, $\delta \Gamma^{\rho}_{
~\mu\nu}(t,\vec{k})$ stands for the spatial Fourier transform of the
perturbed affine connection whose position-space result is,
\begin{equation}
\delta \widetilde{\Gamma}^{\rho}_{~\mu\nu}(t,\vec{x}) \equiv \int \!\! 
\frac{d^3k}{(2\pi)^3} \, e^{i \vec{k} \cdot \vec{x}} \, \delta 
\Gamma^{\rho}_{~\mu\nu}(t,\vec{k}) \; .
\end{equation}
To economize on space we will typically suppress the arguments $t$
and $\vec{k}$.  

The $3+1$ decomposition of the perturbed affine connection is,
\begin{eqnarray}
\delta \Gamma^{0}_{~00} = \dot{\Psi} \quad , \quad
\delta \Gamma^{0}_{~0i} = i k_i \Psi \quad , \quad 
\delta \Gamma^{0}_{~ij} = a^2 \delta_{ij} \Bigl[2 H(\Phi \!-\! \Psi)
\!+\! \dot{\Phi}\Bigr] \; , \label{affine1} \\
\delta \Gamma^{i}_{~00} = \frac{i k_i}{a^2} \Psi \quad , \quad
\delta \Gamma^{i}_{~0j} = \delta_{ij} \dot{\Phi} \quad , \quad 
\delta \Gamma^{i}_{~jk} = i \Bigl[\delta_{ij} k_k \!+\! \delta_{ik} k_j 
\!-\! \delta_{jk} k_i \Bigr] \Phi \; . \label{affine2}
\end{eqnarray}
The corresponding $3+1$ decomposition of the perturbed Ricci tensor is,
\begin{eqnarray}
\delta R_{00} & \!\!\!\!=\!\!\!\! & -6 H \dot{\Phi} \!+\! 3 H \dot{\Psi} 
\!-\! 3 \ddot{\Phi} - \frac{k^2}{a^2} \, \Psi \; , \label{R00} \\
\delta R_{0i} & \!\!\!\!=\!\!\!\! & 2 i k_i \Bigl[H \Psi \!-\! \dot{\Phi}
\Bigr] \; , \label{R0i} \\
\delta R_{ij} & \!\!\!\!=\!\!\!\! & a^2 \delta_{ij} \Bigl[2 (\dot{H} \!+\! 
3 H^2)(\Phi \!-\! \Psi) \!+\! 6 H \dot{\Phi} \!-\! H \dot{\Psi} \!+\! \ddot{\Phi} 
\!+\! \frac{k^2}{a^2} \Phi \Bigr] \!+\! k_i k_j (\Phi \!+\! \Psi) \; . \qquad
\label{Rij} 
\end{eqnarray} 
The associated perturbed Ricci scalar is,
\begin{equation}
\delta R = -12 ( \dot{H} \!+\! 2 H^2) \Psi \!+\! 24 H \dot{\Phi} \!-\! 
6 H \dot{\Psi} \!+\! 6 \ddot{\Phi} \!+\! \frac{k^2}{a^2} \Bigl(4 \Phi \!+\!
2 \Psi\Bigr) \; .
\end{equation}
And $3+1$ decomposing the perturbed Einstein tensor gives,
\begin{eqnarray}
\delta G_{00} & = & 6 H \dot{\Phi} + 2 \frac{k^2}{a^2} \Phi \; , \label{delG00} \\
\delta G_{0i} & = & 2 i k_i \Bigl[H \Psi \!-\! \dot{\Phi}\Bigr] \; , \label{delG0i} \\
\delta G_{ij} & = & a^2 \delta_{ij} \Bigl[-2 (2\dot{H} \!+\!
3 H^2) (\Phi \!-\! \Psi) \nonumber \\
& & \hspace{2.5cm} - 6 H \dot{\Phi} \!+\! 2 H \dot{\Psi} \!-\! 2 \ddot{\Phi} \!-\! 
\frac{k^2}{a^2} (\Phi \!+\! \Psi)\Bigr] + k_i k_j (\Phi \!+\! \Psi) \; . \qquad 
\label{delGij}
\end{eqnarray}

We must also consider covariant derivatives of scalars, vectors and tensors whose 
background plus perturbed forms are,
\begin{eqnarray}
S(t,\vec{x}) & = & \overline{S}(t) + \delta S(t) e^{i \vec{k} \cdot \vec{x}} 
\; , \label{scalar} \\ 
V^{\mu}(t,\vec{x}) & = & \overline{V}(t) \delta^{\mu}_{~0} + \delta V^{\mu}(t) 
e^{i \vec{k} \cdot \vec{x}} \; , \label{vector} \\
\tau_{\mu\nu}(t,\vec{x}) & = & \overline{\tau}(t) \delta^0_{~\mu} \delta^0_{~\nu} + 
\delta \tau_{\mu\nu}(t) e^{i \vec{k} \cdot \vec{x}} \; . \label{tensor} 
\end{eqnarray}
Of course the scalar results are simplest,
\begin{eqnarray}
\delta (\partial_0 S) & = & \delta \dot{S} \qquad , \qquad 
\delta (\partial_i S) = i k_i \delta S \; , \\
\delta (D_0 D_0 S) & = & -\dot{\Psi} \dot{\overline{S}} + \delta \ddot{S} \; , 
\label{d0d0S} \\
\delta (D_0 D_i S) & = &  i k_i \Bigl[-\Psi \dot{\overline{S}} +
(\partial_t \!-\! H) \delta S\Bigr] = \delta (D_i D_0 S) \; , \label{d0diS} \\
\delta (D_i D_j S) & = & -a^2 \delta_{ij} \Bigl[2 H (\Phi \!-\! \Psi) \!+\! 
\dot{\Phi}\Bigr] \dot{\overline{S}} -a^2 \delta_{ij} H \delta \dot{S} 
-k_i k_j \delta S \; . \label{didjS}
\end{eqnarray}
Combining relations (\ref{d0d0S}), (\ref{didjS}) and background results implies,
\begin{equation}
\delta (\square S) = 2 \Psi \Bigl[ \ddot{\overline{S}} \!+\! 3 H \dot{\overline{S}}
\Bigr] + \Bigl[\dot{\Psi} \!-\! 3 \dot{\Phi}\Bigr] \dot{\overline{S}} -
\Bigl[\partial_t^2 \!+\! 3 H \partial_t \!+\! \frac{k^2}{a^2}\Bigr] \delta S \; .
\label{boxS}
\end{equation}

Homogeneity and isotropy imply that the spatial part of vector perturbation 
takes the form,
\begin{equation}
\delta V^i \equiv \frac{i k_i}{a^2} \, \delta V \; .
\end{equation}
The various perturbed first covariant derivatives are,
\begin{eqnarray}
\delta (D_0 V^0) = \dot{\Psi} \overline{V} \!+\! \delta \dot{V}^0 
& \!\!\!\!,\!\!\!\! &
\delta (D_i V^0) = i k_i \Bigl[\Psi \overline{V} \!+\! \delta V^0 \!+\! 
H \delta V\Bigr] , \\
\delta (D_0 V^j) = \frac{i k_j}{a^2} \Bigl[\Psi \overline{V} \!+\! 
(\partial_t \!-\! H) \delta V \Bigr] & \!\!\!\!,\!\!\!\! & 
\delta (D_i V^j) = \delta_{ij} \Bigl[ \dot{\Phi} \overline{V} \!+\! H 
\delta V^0\Bigr] \!-\! \frac{k_i k_j}{a^2} \delta V . \quad
\end{eqnarray}
Combining results gives the perturbed divergence,
\begin{equation}
\delta (D_{\mu} V^{\mu}) = \overline{V} (3 \dot{\Phi} \!+\! \dot{\Psi}) + 
(\partial_t \!+\! 3 H) \delta V^0 - \frac{k^2}{a^2} \delta V \; . \label{Vdel}
\end{equation}
It turns out that we do not require the perturbed second covariant derivatives.

Perturbed tensor covariant derivatives can become very complicated so
it is best to restrict both the form of the perturbation and the 
derivatives we need. The point is to perturb the second derivative 
terms on the 3rd line of equation (\ref{graveqns}), so the tensor is
$\tau_{\mu\nu} = \xi u_{\mu} u_{\nu}$. Because the spatial components 
$u_i$ are already first order, we can specialize to a perturbation of
the form,
\begin{equation}
\delta \tau_{00} \quad , \quad \delta \tau_{0i} = \delta \tau_{i0} = 
i k_i \delta \tau_0 \quad , \quad \delta \tau_{ij} = 0 \; . \label{deltau}
\end{equation}
And the perturbed derivative we require is,
\begin{equation}
\mathcal{T}_{\mu\nu} \equiv \delta \Bigl( -\square \tau_{\mu\nu} - g_{\mu\nu}
D_{\alpha} D_{\beta} \tau^{\alpha\beta} + 2 D_{\alpha} D_{(\mu} 
\tau_{\nu)}^{~~\alpha} \Bigr) \; .
\end{equation}
Homogeneity and isotropy restrict the form of $\mathcal{T}_{\mu\nu}$,
\begin{equation}
\mathcal{T}_{00} \quad , \quad \mathcal{T}_{0i} = \mathcal{T}_{i0} = i k_i 
\mathcal{T}_0 \quad , \quad \mathcal{T}_{ij} = a^2 \delta_{ij} \mathcal{T} 
- k_i k_j \Delta \mathcal{T} \; . \label{4Ts} 
\end{equation}
In Appendix A we derive the following results for the four components of
relation (\ref{4Ts}),
\begin{eqnarray}
\lefteqn{\mathcal{T}_{00} = 3 (\partial_t \!+\! 6 H) (\overline{\tau} \dot{\Phi}) - 
3 H \overline{\tau} \dot{\Psi} - 6 (\partial_t \!+\! 3 H) (H \overline{\tau} \Psi) -
\frac{k^2}{a^2} \, \overline{\tau} \Psi } \nonumber \\
& & \hspace{4.5cm} + 3 (\partial_t \!+\! 3 H) (H \delta \tau_{00}) + \frac{k^2}{a^2} 
\Bigl[\delta \tau_{00} + 4 H \delta \tau_{0}\Bigr] \; , \qquad 
\label{T00} \\
\lefteqn{\mathcal{T}_0 = 2 \overline{\tau} (\partial_t \!+\! H) \Psi + 
3 \dot{\overline{\tau}} \Psi - (\partial_t \!+\! H) \delta \tau_{00} + 2 (\dot{H} 
\!+\! 3 H^2) \delta \tau_{0} \; , } \label{T0} \\
\lefteqn{ \mathcal{T} = -\Bigl[ \overline{\tau} \ddot{\Phi} \!+\! 4 \dot{\overline{\tau}} 
\dot{\Phi} \!+\! 6 H \overline{\tau} \dot{\Phi}\Bigr] + \Bigl[2 \overline{\tau} \ddot{\Psi}
\!+\! 5 \dot{\overline{\tau}} \dot{\Psi} \!+\! 9 H \overline{\tau} \dot{\Psi}\Bigr] 
+ \frac{k^2}{a^2} \, \overline{\tau} \Psi } \nonumber \\
& & \hspace{0cm} - 2 (\Phi \!-\! 2 \Psi) (\partial_t \!+\! 3 H) (\partial_t \!+\! H) 
\overline{\tau} - (\partial_t \!+\! 3 H) (\partial_t \!+\! H) \delta \tau_{00} - 
\frac{2 k^2}{a^2} \, \partial_t \delta \tau_0 \; , \qquad \label{T} \\
\lefteqn{\Delta \mathcal{T} = -2 (\partial_t \!+\! H) \delta \tau_0 \; .} \label{DT}
\end{eqnarray}

\subsection{Perturbed Auxiliary Scalar Equations}

Let us first note from equation (\ref{udef}) that the components of the 
perturbed 4-velocity are,
\begin{equation}
\delta u^0 = -\Psi \qquad , \qquad \delta u^i = \frac{-i k_i \delta \chi}{a^2 
\dot{\chi}_0} \; . \label{delu}
\end{equation}
Now apply relations (\ref{R00}) and (\ref{boxS}) to (\ref{scalars1}) to infer
the equation for $\delta \phi$,
\begin{equation}
\Bigl[ \partial_t^2 \!+\! 3 H \partial_t \!+\! \frac{k^2}{a^2}\Bigr] \delta \phi
= \Bigl[6 \partial_t \!+\! 12 H \!-\! 3 \dot{\phi}_0\Bigr] \dot{\Phi} +
\Bigl[-6 H \partial_t \!+\! \dot{\phi}_0 \partial_t \!+\! \frac{2 k^2}{a^2}\Bigr]
\Psi \; . \label{delphi}
\end{equation}
Note that this gives us the perturbation in $Z$,
\begin{equation}
\delta Z = 2 Z_0 \Bigl[ \frac{\delta \dot{\phi}}{\dot{\phi}_0} - \Psi\Bigr] \; .
\label{delZ}
\end{equation}
The equation for $\delta \chi$ requires only relations (\ref{boxS}) and 
(\ref{scalars1}),
\begin{equation}
\Bigl[ \partial_t^2 \!+\! 3 H \partial_t \!+\! \frac{k^2}{a^2}\Bigr] \delta \chi
= -3 \dot{\chi}_0 \dot{\Phi} + \Bigl[\dot{\chi}_0 \partial_t \!+\! 2\Bigr]
\Psi \; . \label{delchi}
\end{equation}
Using relations (\ref{boxS}) and (\ref{Vdel}) in (\ref{scalars2}) gives the
equation for $\delta \xi$,
\begin{equation}
\Bigl[ \partial_t^2 \!+\! 3 H \partial_t \!+\! \frac{k^2}{a^2}\Bigr] \delta \xi
= 2 (\partial_t \!+\! 3 H) \Bigl[ f_y'(Z_0) \delta \dot{\phi} \!+\! f_y''(Z_0)
\dot{\phi}_0 \delta Z\Bigr] + \frac{2 k^2}{a^2} f_y'(Z_0) \delta \phi \; .
\label{delxi}
\end{equation}
And our equation for $\delta \psi$ comes from using relations (\ref{R0i}),
(\ref{boxS}) and (\ref{Vdel}) in (\ref{scalars2}),
\begin{equation}
\Bigl[ \partial_t^2 \!+\! 3 H \partial_t \!+\! \frac{k^2}{a^2}\Bigr] \delta \psi
= \frac{k^2}{a^2} \frac{8 \xi_0}{\dot{\chi}_0} \Bigl[ 
\frac{\dot{H} \delta \chi}{\dot{\chi}_0} \!+\! H \Psi \!-\! \dot{\Phi} \Bigr] 
\; . \label{delpsi}
\end{equation}

\subsection{Perturbed Gravitational Field Equations}

We employ the following notation to express the modified Einstein equation,
\begin{equation}
\mathcal{E}_{\mu\nu} \equiv G_{\mu\nu} + \mathcal{G}_{\mu\nu} = 8\pi G
T_{\mu\nu} \; . \label{modEin}
\end{equation}
Here $G_{\mu\nu}$ the usual Einstein tensor, $\mathcal{G}_{\mu\nu}$ is the MOND
correction to it given in equation (\ref{graveqns}) and $T_{\mu\nu}$ is the stress
tensor without dark matter. Relations (\ref{FLRW00}) and (\ref{FLRWij}) give the
nonzero components of (\ref{modEin}) when the geometry is specialized to a 
cosmological background. We denote the first order perturbations of 
$\mathcal{E}_{\mu\nu}$ and $T_{\mu\nu}$ as,
\begin{eqnarray}
\delta \mathcal{E}_{00} \qquad , \qquad \delta \mathcal{E}_{0i} = \delta \mathcal{E}_{i0}
= i k_i \delta \mathcal{E}_0 \qquad , \qquad \delta \mathcal{E}_{ij} = a^2 \delta_{ij}
\delta \mathcal{E} - k_i k_j \Delta \mathcal{E} \; , \\
\delta T_{00} = \delta \rho \qquad , \qquad \delta T_{0i} = \delta T_{i0} = i k_i 
\Delta \rho \qquad , \qquad \delta T_{ij} = a^2 \delta_{ij} \delta T \; . 
\end{eqnarray}
The various components of the perturbed stress tensor are related by the two 
conservation equations,
\begin{eqnarray} 
0 & \!\!\!\!=\!\!\!\! & -\delta \dot{\rho} - 3 H (\delta \rho \!+\! \delta T) + 
\frac{k^2}{a^2} \Delta \rho - 3 (\rho_0 \!+\! p_0) \dot{\Phi} + 2 \rho_0 \dot{\Psi}
+ 6 H p_0 (\Phi \!-\! \Psi) \; , \qquad \\
0 & \!\!\!\!=\!\!\!\! & -(\partial_t \!+\! 3 H) \Delta \rho + \delta T - 2 p_0 
\Phi + (\rho_0 \!+\! p_0) \Psi \; .
\end{eqnarray}
Similar relations hold for the components of $\delta \mathcal{E}_{\mu\nu}$. 
Because $\delta \mathcal{E}$ follows from conservation we will not report it.

Relations (\ref{4Ts}-\ref{DT}) give the perturbed second covariant derivatives 
of a tensor $\tau_{\mu\nu} = \xi u_{\mu} u_{\nu}$. From expression (\ref{delu}) 
we see that the components (\ref{deltau}) of this tensor are,
\begin{equation}
\overline{\tau} = \xi_0 \qquad , \qquad \delta \tau_{00} = 2 \Psi \xi_0 + 
\delta \xi \qquad , \qquad \delta \tau_0 = \frac{\xi_0 \delta \chi}{\dot{\chi}_0}
\; . \label{taucomps}
\end{equation}
From the $00$ component of (\ref{graveqns}), with relations (\ref{R00}), 
(\ref{delG00}), (\ref{T00}) and (\ref{taucomps}) we infer the perturbed $00$ 
equation,
\begin{eqnarray}
\lefteqn{\Bigl[6 H \!+\! 3 \dot{\xi}_0 \!+\! 12 H \xi_0\Bigr] \dot{\Phi} + a_0^2
f_y(Z_0) \Psi + a_0^2 \Bigl[ \frac{f_y'(Z_0)}{2} \!-\! Z_0 f_y''(Z_0)\Bigr] \delta Z
+ \frac{\dot{\xi}_0 \delta \dot{\phi} \!-\! \dot{\phi}_0 \delta \dot{\xi}}{2} } 
\nonumber \\
& & \hspace{0cm} + \Bigl[ 3 H \partial_t \!+\! 6 H^2 \Bigr] \delta \xi \!-\! 
\Bigl[ \dot{\chi}_0 \partial_t \!-\! 1\Bigr] \frac{\delta \psi}{2} \!+\! 
\frac{k^2}{a^2} \Biggl[ 2 \Phi \!+\! \delta \xi \!+\! 
\frac{4 H \xi_0 \delta \chi}{\dot{\chi}_0} \Biggr] = 8 \pi G \delta \rho \; . 
\qquad \label{delE00}
\end{eqnarray}
The perturbed $0i$ components follow from using (\ref{R0i}), (\ref{delG0i}), 
(\ref{T0}) and (\ref{taucomps}) in expression (\ref{graveqns}), 
\begin{equation}
\Bigl(2 H \!+\! 4 H \xi_0 \!+\! \dot{\xi}_0\Bigr) \Psi - \Bigl(2 \!+\! 4 \xi_0\Bigr)
\dot{\Phi} - \Bigl(\partial_t \!+\! H \!+\! \frac12 \dot{\phi}_0\Bigr) \delta \xi 
-\frac12 \dot{\chi}_0 \delta \psi = 8 \pi G \Delta \rho \; . \label{delE0i}
\end{equation}
Finally, expressions (\ref{delGij}) and (\ref{DT}) give us the gravitational 
slip equation,
\begin{equation}
\Phi + \Psi + 2 (\partial_t \!+\! H) \Bigl[ \frac{\xi_0 \delta \chi}{\dot{\chi}_0}
\Bigr] = 0 \; . \label{slip}
\end{equation}

\subsection{The Sub-Horizon Regime}

The perturbed equations of the previous two sub-sections simplify dramatically
in the sub-horizon regime of $k \gg H a$. In this regime time derivatives of
the fields are also negligible so the various auxiliary scalar equations 
(\ref{delphi}), (\ref{delchi}), (\ref{delxi}) and (\ref{delpsi}) can be used
to express the scalar perturbations in terms of the two gravitational
potentials,
\begin{equation}
k \gg H a \Longrightarrow \delta \phi = 2 \Psi \; , \; \delta \chi = 0 \; , \;
\delta \xi = 4 f_y'(Z_0) \Psi \; , \; \delta \psi = \frac{8 \xi_0}{\dot{\chi}_0}
\Bigl[ H \Psi \!-\! \dot{\Phi}\Bigr] \; . \label{subscalars}
\end{equation}
Making the same approximations in the $00$ component (\ref{delE00}) of the
perturbed gravitational field equations implies,
\begin{equation}
k \gg H a \qquad \Longrightarrow \qquad \frac{k^2}{a^2} \Bigl[ 2\Phi + 4 f_y'(Z_0) 
\Psi\Bigr] = 8\pi G \delta \rho \; . \label{subE00}
\end{equation}
The $0i$ equation (\ref{delE0i}) reduces to,
\begin{equation}
k \gg H a \qquad \Longrightarrow \qquad 2 H \Psi - 2 \dot{\Phi} - 4 
(\partial_t \!+\! H) \Bigl[ f_y'(Z_0) \Psi\Bigr] = 8 \pi G \Delta \rho \; . 
\label{subE0i}
\end{equation}
And the gravitational slip equation (\ref{slip}) reduces to that of unmodified
general relativity,
\begin{equation}
k \gg H a \qquad \Longrightarrow \qquad \Phi + \Psi = 0 \; . \label{subslip}
\end{equation}

Relations (\ref{subE00}) and (\ref{subslip}) are fatal because the first term 
in the square brackets of (\ref{subE00}) --- $2\Phi$ --- is the contribution from 
unmodified general relativity, whereas the second term --- $4 f_y'(Z_0) \Psi =
-4 f_y'(Z_0) \Phi$ using (\ref{subslip}) --- is the MOND correction. The right 
hand side of (\ref{subE00}) is the perturbed energy density {\it without} dark 
matter. For this model to be viable the MOND correction must largely cancel the 
factor of $2 \Phi$ from general relativity, so that one gets the same gravitational 
response from the much smaller source. However, a glance at Fig.~\ref{fyprime} 
reveals that the function $f_y'(Z_0) \ltwid 0.015$ is much too small, and actually
{\it strengthens} the general relativistic result for $z \ltwid 6$. Hence the
gravitational response to matter perturbations is much too weak.

\section{Discussion}

In this paper we have considered structure formation in a nonlocal,
metric-based realization of MOND \cite{Deffayet:2011sk,Deffayet:2014lba}.
This model involves augmenting the gravitational action by an algebraic
function $f_y(Z)$ of a nonlocal invariant. The function $f_y(Z)$ is constrained
for positive $Z$ by the Tully-Fisher relation (for $0 < Z \ll 1$) and by the
need to leave solar system results undisturbed (for $1 \ltwid Z$). Cosmology
corresponds to $Z < 0$ and the function was chosen to reproduce the $\Lambda$CDM 
expansion history (until very late times) without dark matter \cite{Kim:2016nnd}. 

Section 2 was devoted to describing the model in general, and as specialized to 
static, spherically symmetric geometries and to homogeneous and isotropic 
geometries. In section 3 we derived the equations --- (\ref{delphi}-\ref{delpsi})
for the auxiliary scalars and (\ref{delE00}-\ref{slip}) for the gravitational
potentials --- governing first order scalar perturbations around the cosmological 
background. Specializing those equations to the sub-horizon regime leads to vastly 
simpler equations --- (\ref{subscalars}) for the auxiliary scalars and 
(\ref{subE00}-\ref{subslip}) for the gravitational potentials. These simplified 
equations show that the MOND corrections cannot possibly make up for the absence 
of dark matter in structure formation. Hence this particular model is falsified.

It is interesting to contrast our negative result with what happens in a nonlocal 
cosmology model that was proposed explain cosmic acceleration without dark 
energy \cite{Deser:2007jk}. This model cannot reproduce MOND \cite{Arraut:2013qra}
but its free function $f(X)$ (of a different nonlocal invariant) can be adjusted 
to enforce the $\Lambda$CDM expansion history \cite{Deffayet:2009ca} without a 
cosmological constant. However, when linearized scalar perturbations are studied 
in nonlocal cosmology \cite{Park:2012cp,Dodelson:2013sma}, the ``extra'' terms 
which are not present in general relativity turn out to be significant,
and actually cause the model to agree better with data \cite{Nersisyan:2017mgj,
Park:2017zls}. For both nonlocal cosmology and nonlocal MOND, the ``extra''
terms are proportional to the derivative of the free function --- $f'(X_0)$ for
nonlocal cosmology and $f'_y(Z_0)$ for nonlocal MOND. The key distinction between
the two models is that $f'(X_0)$ is significant for nonlocal cosmology, whereas 
$f'_y(Z_0)$ is nearly zero for nonlocal MOND. That seems to be reason why the
results are so very different. 

That $f_y'(Z_0)$ must be small for very large redshift follows from the
asymptotic relation (\ref{asympfyfy'}), which could be written,
\begin{equation}
f_y'(Z_0) \longrightarrow \frac1{44} \times \frac{\Omega_c}{\Omega_m} \times 
\Bigl(\frac{1 + z_{\rm eq}}{1 + z}\Bigr) \simeq 0.019 \times \Bigl(
\frac{3300}{1 + z}\Bigr) \; . \label{badform}
\end{equation}
However, this relation only pertains for $z > z_{eq}$, during the radiation 
dominated phase when matter is unimportant. There is no simple way to understand
the crucial behavior of $f_y'(Z_0)$ for smaller redshift, which is shown in
Fig.~\ref{fyprime}. The fact that $f_y(Z_0)$ approaches the limiting form 
(\ref{badform}) from {\it below}, and actually goes to {\it negative infinity}
at $z = z_*$, seems to follow from our decision to solve (\ref{feqn}) with the
initial condition $f(z_*) = 0$. In retrospect, it might be more reasonable to
impose the condition $f'(z_*) = 0$, which would keep $f_y'(Z_0)$ finite at 
$z = z_*$, and might result in $f_y'(Z_0)$ approaching the limiting form 
(\ref{badform}) {\it from above}. This would be the simplest fix because it
would leave the equations for linearized perturbations unchanged when written
in terms of the generic background quantities, changing only the numerical 
values of those background quantities.

Two more complicated fixes are also conceivable. The first would be to involve 
another invariant as suggested in the original proposal \cite{Deffayet:2011sk} 
for nonlocal MOND. This extra invariant is not required to reproduce either
the Tully-Fisher relation, or (most of) the $\Lambda$CDM expansion history, but
perhaps it plays a crucial role in structure formation.

The second conceivable extension of the model would be to make $a_0$ dynamical.
The numerical coincidence that $a_0$ is about $c H_0/2\pi$ has led many to
suspect that $a_0$ is not actually a new constant but rather a functional of
the geometry which is always close to $c$ times the Hubble parameter
\cite{Bekenstein:2008pc,Milgrom:2008cs}. The idea would be to keep cosmology 
in the deep MOND regime and NOT use the function $f_y(Z)$ to enforce the 
$\Lambda$CDM expansion history.

\section{Appendix: Perturbed Tensor Derivatives}

The purpose of this appendix is to derive relations (\ref{T00}-\ref{DT}). Our
strategy is to construct the final answer in three steps:
\begin{enumerate}
\item{Expand $\mathcal{T}_{\mu\nu}$ in terms of $\delta g_{\mu\nu}$ and 
$\delta (D_{\rho} D_{\sigma} \tau_{\mu\nu})$;}
\item{Expand $\delta (D_{\rho} D_{\sigma} \tau_{\mu\nu})$ in terms of $\delta 
\Gamma^{\rho}_{~\mu\nu}$ and $\delta (D_{\rho} \tau_{\mu\nu})$; and}
\item{Expand $\delta (D_{\rho} \tau_{\mu\nu})$ in terms of $\delta \Gamma^{\rho}_{~\mu\nu}$
and $\delta \tau_{\mu\nu}$.}
\end{enumerate}
Because many cancellations occur at each level we combine terms in $\mathcal{T}_{00}$,
$\mathcal{T}_{0i}$ and $\mathcal{T}_{ij}$ after each expansion, before proceeding to the
next step. The nonzero components of the background quantities we require are,
\begin{eqnarray}
\overline{\Gamma}^0_{~ij} = H a^2 \delta_{ij} & , & \overline{\Gamma}^i_{~0j} = 
H \delta_{ij} \; , \\
\overline{D_0 \tau_{00}} = \dot{\overline{\tau}} & , & \overline{D_i \tau_{0j}} = -H
\overline{\tau} a^2 \delta_{ij} \; , \\
\overline{D_0 D_0 \tau_{00}} = \ddot{\overline{\tau}} & , & \overline{D_i D_j \tau_{00}} =
-H (\partial_t \!-\! 2 H) \overline{\tau} a^2 \delta_{ij} \; , \\
\overline{D_0 D_i \tau_{0j}} = -\partial_t (H \overline{\tau}) a^2 \delta_{ij} & , & 
\overline{D_i D_0 \tau_{0j}} = -H (\partial_t \!-\! H) \overline{\tau} a^2 \delta_{ij} 
\; , 
\end{eqnarray}
and
\begin{equation}
\overline{D_i D_j \tau_{k\ell}} = H^2 \overline{\tau} a^4 (\delta_{ik} \delta_{j\ell} \!+\!
\delta_{i\ell} \delta_{jk}) \; .
\end{equation} 
 
The first step expansions are:
\begin{eqnarray}
\lefteqn{-\delta (\square \tau_{\mu\nu}) = -2 \Psi \overline{D_0 D_0 \tau_{\mu\nu}} +
\frac{2 \Phi}{a^2} \, \overline{D_k D_k \tau_{\mu\nu}} } \nonumber \\
& & \hspace{5.5cm} + \delta (D_0 D_0 \tau_{\mu\nu}) -\frac1{a^2} \, \delta 
(D_k D_k \tau_{\mu\nu}) \; , \qquad \\
\lefteqn{-\delta (g_{\mu\nu} D_{\alpha} D_{\beta} \tau^{\alpha\beta}) = \delta g_{\mu\nu}
\Biggl[-\overline{D_{0} D_{0} \tau_{00}} \!+\! \frac{( \overline{D_0 D_k \tau_{0k}} \!+\! 
\overline{D_{k} D_0 \tau_{k0}})}{a^2} \!-\! \frac{\overline{D_k D_{\ell} \tau_{k\ell}}}{a^4} 
\Biggr]} \nonumber \\
& & \hspace{-.5cm} + \overline{g_{\mu\nu}} \Biggl[ 4\Psi \overline{D_0 D_0 \tau_{00}} 
- \frac{2 (\Psi \!+\! \Phi)}{a^2} \Bigl( \overline{ D_0 D_k \tau_{0k}} \!+\! 
\overline{D_k D_0 \tau_{k0}} \Bigr) + \frac{4 \Phi}{a^4} \overline{D_k D_{\ell} \tau_{k\ell}} 
\nonumber \\
& & \hspace{.5cm} - \delta (D_0 D_0 \tau_{00}) + \frac{1}{a^2} \delta \Bigl(D_0 D_k \tau_{0k} 
\!+\! D_k D_0 \tau_{k0}\Bigr) - \frac1{a^4} \delta(D_k D_{\ell} \tau_{k\ell}) \Biggr] 
, \qquad \\
\lefteqn{\delta ( 2 D_{\alpha} D_{(\mu} \tau_{\nu)}^{~~\alpha}) = 4\Psi \overline{D_0 D_{(\mu} 
\tau_{\nu) 0}} - \frac{4 \Phi}{a^2} \, \overline{D_k D_{(\mu} \tau_{\nu) k}} } \nonumber \\
& & \hspace{5cm} -2 \delta (D_0 D_{(\mu} \tau_{\nu) 0}) + \frac2{a^2} \, \delta 
(D_k D_{(\mu} \tau_{\nu) k}) \; . \qquad 
\end{eqnarray}
The $3+1$ totals after the 1st step are,
\begin{eqnarray}
\mathcal{T}_{00} & = & -6 H (\partial_t \!+\! 7 H) (H \overline{\tau}) \, \Phi
+ 24 H^2 \overline{\tau} \, \Psi \nonumber \\
& & + \frac1{a^2} \delta \Bigl( -D_k D_k \tau_{00} \!-\! D_0 D_k \tau_{0k} \!+\! 
D_k D_0 \tau_{0k}\Bigr) + \frac{ \delta (D_k D_{\ell} \tau_{k\ell})}{a^4} \; , \label{T001} \\
\mathcal{T}_{0i} & = & -\delta (D_0 D_i \tau_{00}) + \frac1{a^2} \delta \Bigl( -D_k D_k \tau_{0i}
\!+\! D_k D_0 \tau_{ik} \!+\! D_k D_i \tau_{0k}\Bigr) \; , \label{T0i1} \\
\mathcal{T}_{ij} & = & a^2 \delta_{ij} \Biggl[ \Bigl( -2 \ddot{\overline{\tau}} \!+\! 12 H^2 
\overline{\tau}\Bigr) \Phi + \Bigl( 4 \ddot{\overline{\tau}} \!+\! 8 H \dot{\overline{\tau}} \!+\!
(2 \dot{H} \!-\! 6 H^2) \overline{\tau}\Bigr) \Psi \nonumber \\
& & -\delta (D_0 D_0 \tau_{00}) + \frac1{a^2} \delta \Bigl( D_0 D_k \tau_{0k} \!+\! D_k D_0 \tau_{0k}
\Bigr) - \frac1{a^4} \delta( D_k D_{\ell} \tau_{k\ell}) \Biggr] \nonumber \\
& & \hspace{.3cm} + \delta \Bigl( D_0 D_0 \tau_{ij} \!-\! 2 D_0 D_{(i} \tau_{j) 0}\Bigr) + 
\frac1{a^2} \delta \Bigl( -D_k D_k \tau_{ij} \!+\! 2 D_{k} D_{(i} \tau_{j) k}\Bigr) \; . 
\label{Tij1} \qquad 
\end{eqnarray}
Note that only the terms on the last line of (\ref{Tij1}) can possibly contribute to
the $k_i k_j$ part of $\mathcal{T}_{ij}$.

Of course it is only necessary to implement the second step for the $\delta (D_{\rho}
D_{\sigma} \tau_{\mu\nu})$ parts of expressions (\ref{T001}-\ref{Tij1}). The key expansion 
is,
\begin{eqnarray}
\lefteqn{ \delta (D_{\rho} D_{\sigma} \tau_{\mu\nu}) = -\delta \Gamma^{\alpha}_{~\rho\sigma} \,
\overline{D_{\alpha} \tau_{\mu\nu}} - \delta \Gamma^{\alpha}_{~ \rho\mu} \, \overline{D_{\sigma}
\tau_{\alpha \nu}} - \delta \Gamma^{\alpha}_{~\rho\nu} \, \overline{D_{\sigma} \tau_{\mu\alpha}} }
\nonumber \\
& & \hspace{.5cm} + \partial_{\rho} \delta (D_{\sigma} \tau_{\mu\nu}) - 
\overline{\Gamma}^{\alpha}_{~\rho\sigma} \, \delta (D_{\alpha} \tau_{\mu\nu}) - 
\overline{\Gamma}^{\alpha}_{~\rho\mu} \, \delta (D_{\sigma} \tau_{\alpha\nu}) - 
\overline{\Gamma}^{\alpha}_{~\rho\nu} \, \delta (D_{\sigma} \tau_{\mu\alpha}) \; . \qquad 
\end{eqnarray}
The $\delta (D_{\rho} D_{\sigma} \tau_{\mu\nu})$ part of $\mathcal{T}_{00}$ from 
expression (\ref{T001}) gives,
\begin{eqnarray}
\lefteqn{\frac1{a^2} \delta \Bigl( -D_k D_k \tau_{00} \!-\! D_0 D_k \tau_{0k} \!+\! 
D_k D_0 \tau_{0k}\Bigr) + \frac{ \delta (D_k D_{\ell} \tau_{k\ell})}{a^4} } \nonumber \\
& & \hspace{.5cm} = 3 H \overline{\tau} \Bigl[ (\dot{\Phi} - \dot{\Psi}) + 8 H (\Phi - \Psi)
\Bigr] -\frac1{a^2} (\partial_0 \!+\! H) \delta (D_k \tau_{0k}) \nonumber \\
& & \hspace{2cm} -\frac{2 H}{a^2} \, \delta (D_0 \tau_{kk}) + \frac{i k_k}{a^2} \, \delta 
\Bigl(D_0 \tau_{0k} - D_k \tau_{00}\Bigr) + \frac{i k_k}{a^4} \, \delta (D_{\ell} \tau_{k\ell}) \; . 
\qquad \label{T002}
\end{eqnarray}
The step 1 reduction of $\mathcal{T}_{0i}$ in expression (\ref{T0i1}) contains only
$\delta (D_{\rho} D_{\sigma} \tau_{\mu\nu})$ terms,
\begin{eqnarray}
\lefteqn{\mathcal{T}_{0i} = ik_i (\dot{\overline{\tau}} \!-\! 2 H \overline{\tau}) \Psi - 
(\partial_0 \!+\! H) \delta (D_i \tau_{00}) } \nonumber \\
& & \hspace{.5cm} + \frac{ik_k}{a^2} \, \delta \Bigl(- D_k \tau_{0i} \!+\! D_0 \tau_{ik} \!+\! 
D_i \tau_{0k}\Bigr) - 2 H \delta (D_0 \tau_{0i}) -\frac{H}{a^2} \, \delta (D_i \tau_{kk}) \; . 
\qquad \label{T0i2}
\end{eqnarray}
The step 1 reduction of $\mathcal{T}_{ij}$ in expression (\ref{Tij1}) contained 
some terms which are already proportional to $a^2 \delta_{ij}$,
\begin{eqnarray}
\lefteqn{-\delta (D_0 D_0 \tau_{00}) + \frac1{a^2} \delta \Bigl( D_0 D_k \tau_{0k} \!+\! 
D_k D_0 \tau_{0k} \Bigr) - \frac1{a^4} \delta( D_k D_{\ell} \tau_{k\ell}) } \nonumber \\
& & \hspace{.5cm} = -3 (\dot{\overline{\tau}} \!+\! H \overline{\tau}) (\dot{\Phi} \!-\!
\dot{\Psi}) - 6 H (\dot{\overline{\tau}} \!+\! 4 H \overline{\tau}) (\Phi \!-\! \Psi) 
- (\partial_0 \!+\! 3 H) \delta (D_0 \tau_{00}) \nonumber \\
& & \hspace{2.5cm} + \frac1{a^2} \, (\partial_0 \!+\! H) \delta (D_k \tau_{0k}) + 
\frac{i k_k}{a^2} \, \delta (D_0 \tau_{0k}) - \frac{i k_k}{a^4} \delta (D_{\ell} \tau_{k\ell}) 
\; . \qquad \label{Tij2a}
\end{eqnarray}
The other $\delta (D_{\rho} D_{\sigma} \tau_{\mu\nu})$ terms in expression (\ref{Tij1})
are,
\begin{eqnarray}
\lefteqn{ \delta \Bigl( D_0 D_0 \tau_{ij} \!-\! 2 D_0 D_{(i} \tau_{j) 0}\Bigr) + \frac1{a^2} 
\delta \Bigl( -D_k D_k \tau_{ij} \!+\! 2 D_{k} D_{(i} \tau_{j) k}\Bigr) } \nonumber \\
& & \hspace{1cm} = 2 H \overline{\tau} a^2 \delta_{ij} \Bigl[ (\dot{\Phi} \!-\! \dot{\Phi}) 
+ 6 H (\Phi \!-\! \Psi)\Bigr] + (\partial_0 \!-\! H) \delta (D_0 \tau_{ij}) \nonumber \\
& & \hspace{3.5cm} - 2 (\partial_0 \!+\! H) \delta (D_i \tau_{0j}) + \frac{i k_k}{a^2} \,
\delta \Bigl[- D_k \tau_{ij} + 2 D_{i} \tau_{jk}\Bigr] \; . \qquad \label{Tij2b}
\end{eqnarray}

As before, it is only necessary to implement the 3rd step reduction for those parts of
expressions (\ref{T002}-\ref{Tij2b}) which contain $\delta ( D_{\rho} \tau_{\mu\nu})$. 
The key reduction of step 3 is,
\begin{equation}
\delta (D_{\rho} \tau_{\mu\nu}) = -\delta \Gamma^{\sigma}_{~ \rho\mu} \overline{\tau}_{\sigma\nu}
- \delta \Gamma^{\sigma}_{~\rho\nu} \overline{\tau}_{\mu\sigma} + \partial_{\rho} \delta
\tau_{\mu\nu} - \overline{\Gamma}^{\sigma}_{~\rho\mu} \delta \tau_{\sigma\nu} - 
\overline{\Gamma}^{\sigma}_{~\rho\nu} \delta \tau_{\mu\sigma} \; .
\end{equation}
The step 3 reduction of the $\delta ( D_{\rho} \tau_{\mu\nu})$ terms in (\ref{T002}) is,
\begin{eqnarray}
\lefteqn{-\frac1{a^2} (\partial_0 \!+\! H) \delta (D_k \tau_{0k}) \!-\! \frac{2 H}{a^2} \, 
\delta (D_0 \tau_{kk}) \!+\! \frac{i k_k}{a^2} \, \delta  \Bigl(D_0 \tau_{0k} \!-\! 
D_k \tau_{00}\Bigr) \!+\! \frac{i k_k}{a^4} \, \delta (D_{\ell} \tau_{k\ell}) } \nonumber \\
& & \hspace{2cm} = (\partial_t \!+\! 3 H) \Bigl[3 \overline{\tau} \dot{\Phi} \!+\! 6 H 
\overline{\tau} (\Phi \!-\! \Psi)\Bigr] - \frac{k^2}{a^2} \, \overline{\tau} \Psi \nonumber \\
& & \hspace{4.5cm} + 3 (\partial_t \!+\! 3 H) (H \delta \tau_{00}) + \frac{k^2}{a^2} \Bigl[
\delta \tau_{00} + 4 H \delta \tau_{0}\Bigr] \; . \label{T003} \qquad
\end{eqnarray}
The step 3 reduction of the $\delta ( D_{\rho} \tau_{\mu\nu})$ terms in (\ref{T0i2}) is,
\begin{eqnarray}
\lefteqn{ (\partial_0 \!+\! H) \delta (D_i \tau_{00}) \!-\! \frac{ik_k}{a^2} \, \delta 
\Bigl(D_k \tau_{0i} \!-\! D_0 \tau_{ik} \!-\! D_i \tau_{0k}\Bigr) \!-\! 2 H \delta (D_0 \tau_{0i}) 
\!-\! \frac{H}{a^2} \, \delta (D_i \tau_{kk}) } \nonumber \\
& & \hspace{2cm} = i k_i \Bigl[ 2 (\partial_t \!+\! 2 H) (\overline{\tau} \Psi) -
(\partial_t \!+\! H) \delta \tau_{00} + (2 \dot{H} \!+\! 6 H^2) \delta \tau_{0} \Bigr] \; .
\qquad \label{T0i3}
\end{eqnarray}
The step 3 reduction of the $\delta ( D_{\rho} \tau_{\mu\nu})$ terms in (\ref{Tij2a}) is,
\begin{eqnarray}
\lefteqn{ - (\partial_0 \!+\! 3 H) \delta (D_0 \tau_{00}) \!+\! \frac1{a^2} \, (\partial_0 
\!+\! H) \delta (D_k \tau_{0k}) \!+\! \frac{i k_k}{a^2} \, \delta (D_0 \tau_{0k}) \!-\! 
\frac{i k_k}{a^4} \delta (D_{\ell} \tau_{k\ell}) } \nonumber \\
& & \hspace{3cm} = (\partial_t \!+\! 3 H) \Bigl[-3 \overline{\tau} \dot{\Phi} \!+\! 2
\overline{\tau} \dot{\Psi} - 6 H \overline{\tau} (\Phi \!-\! \Psi)\Bigr] + \frac{k^2}{a^2} 
\, \overline{\tau} \Psi \nonumber \\
& & \hspace{5.5cm} -(\partial_t \!+\! 3 H)^2 \delta \tau_{00} - \frac{2 k^2}{a^2} \, 
(\partial_t \!+\! H) \delta \tau_{0} \; . \qquad \label{Tij3a}
\end{eqnarray}
And the step 3 reduction of the $\delta ( D_{\rho} \tau_{\mu\nu})$ terms in (\ref{Tij2b}) 
is,
\begin{eqnarray}
\lefteqn{ (\partial_0 \!-\! H) \delta (D_0 \tau_{ij}) \!-\! 2 (\partial_0 \!+\! H) 
\delta (D_i \tau_{0j}) \!+\! \frac{i k_k}{a^2} \, \delta \Bigl[- D_k \tau_{ij} \!+\! 
2 D_{i} \tau_{jk}\Bigr] = 2 a^2 \delta_{ij} } \nonumber \\
& & \hspace{-.5cm} \times \! \Biggl[ (\partial_t \!+\! 3 H) \Bigl[ \overline{\tau}
\dot{\Phi} \!+\! 2 H \overline{\tau} (\Phi \!-\! \Psi) \!+\! H \delta \tau_{00}\Bigr] \!+\! 
\frac{k^2 H \delta{\tau}_0}{a^2} \Biggr] \!+\! k_i k_j 2 (\partial_t \!+\! H) \delta \tau_{0} 
\; . \qquad \label{Tij3b}
\end{eqnarray}

Combining expressions (\ref{T001}), (\ref{T002}) and (\ref{T003}) gives our final
result for $\mathcal{T}_{00}$,
\begin{eqnarray}
\lefteqn{\mathcal{T}_{00} = 3 (\partial_t \!+\! 6 H) (\overline{\tau} \dot{\Phi}) - 
3 H \overline{\tau} \dot{\Psi} - 6 (\partial_t \!+\! 3 H) (H \overline{\tau} \Psi) -
\frac{k^2}{a^2} \, \overline{\tau} \Psi } \nonumber \\
& & \hspace{4.5cm} + 3 (\partial_t \!+\! 3 H) (H \delta \tau_{00}) + \frac{k^2}{a^2} \Bigl[
\delta \tau_{00} + 4 H \delta \tau_{0}\Bigr] \; . \qquad \label{T00final}
\end{eqnarray}
Our final result for $\mathcal{T}_0$ comes from expressions (\ref{T0i2}) and (\ref{T0i3}),
\begin{equation}
\mathcal{T}_0 = 2 \overline{\tau} (\partial_t \!+\! H) \Psi + 3 \dot{\overline{\tau}} \Psi -
(\partial_t \!+\! H) \delta \tau_{00} + 2 (\dot{H} \!+\! 3 H^2) \delta \tau_{0} \; .
\label{T0final}
\end{equation}
Combining expressions (\ref{Tij1}), (\ref{Tij2a}), (\ref{Tij2b}), (\ref{Tij3a}) and
(\ref{Tij3b}) gives our result for $\mathcal{T}$,
\begin{eqnarray}
\lefteqn{ \mathcal{T} = -\Bigl[ \overline{\tau} \ddot{\Phi} \!+\! 4 \dot{\overline{\tau}} 
\dot{\Phi} \!+\! 6 H \overline{\tau} \dot{\Phi}\Bigr] + \Bigl[2 \overline{\tau} \ddot{\Psi}
\!+\! 5 \dot{\overline{\tau}} \dot{\Psi} \!+\! 9 H \overline{\tau} \dot{\Psi}\Bigr] 
+ \frac{k^2}{a^2} \, \overline{\tau} \Psi } \nonumber \\
& & \hspace{0cm} - 2 (\Phi \!-\! 2 \Psi) (\partial_t \!+\! 3 H) (\partial_t \!+\! H) 
\overline{\tau} - (\partial_t \!+\! 3 H) (\partial_t \!+\! H) \delta \tau_{00} - 
\frac{2 k^2}{a^2} \, \partial_t \delta \tau_0 \; . \qquad \label{Tfinal} 
\end{eqnarray}
And our result for $\Delta \mathcal{T}$ comes entirely from (\ref{Tij3b}),
\begin{equation}
\Delta \mathcal{T} = -2 (\partial_t \!+\! H) \delta \tau_0 \; . \label{DTfinal}
\end{equation}

The principal complication in expressions (\ref{T00final}) and (\ref{Tfinal})
is their dependence upon the potentials $\Phi$ and $\Psi$, and the perturbation
$\delta \tau_{00}$, and their respective time derivatives. This can actually be 
predicted by transforming the background expressions,
\begin{equation}
\overline{\mathcal{T}}_{00} = 3 (\partial_t \!+\! 3 H) (H \overline{\tau}) \qquad , 
\qquad \overline{\mathcal{T}}_{ij} = -a^2 \delta_{ij} (\partial_t \!+\! 3 H)
(\partial_t \!+\! H) \overline{\tau} \; ,
\end{equation}
by the same transformation which would carry the background geometry into the 
perturbed one assuming the potentials depend only on time,
\begin{equation}
-dt^2 + a^2 d\vec{x} \!\cdot\! d\vec{x} \longrightarrow -\Bigl[1 \!+\! 2 \Psi\Bigr]
dt^2 + \Bigl[1 \!+\! 2 \Phi\Bigr] a^2 d\vec{x} \!\cdot\! d\vec{x} \; ,
\end{equation}
Although we only need the first order deviations, the all-orders relations are,
\begin{eqnarray}
\overline{\mathcal{T}}_{00} \!+\! \mathcal{T}_{00} & \!\!\! = \!\!\! & 3 \sqrt{1 \!+\!
2 \Psi} \Biggl[ \frac{d}{dt} \!+\! 3 H \!+\! \frac{3 \dot{\Phi}}{1 \!+\! 2 \Phi}
\Biggr] \Biggl[ \frac{ (H \!+\! \frac{\dot{\Phi}}{1 \!+\! 2 \Phi}) (\overline{\tau} 
\!+\! \delta \tau_{00})}{(1 \!+\! 2 \Psi)^{\frac32}} \Biggr] , \qquad \\
\overline{\mathcal{T}}_{ij} \!+\! \mathcal{T}_{ij} & \!\!\! = \!\!\! & -
\frac{(1 \!+\! 2 \Phi) a^2}{\sqrt{1 \!+\! 2 \Psi}} \Biggl[ \frac{d}{dt} \!+\! 3 H
\!+\! \frac{3 \dot{\Phi}}{1 \!+\! 2 \Phi} \Biggr] \nonumber \\
& & \hspace{3.5cm} \times \frac1{\sqrt{1 \!+\! 2 \Psi}} \Biggl[ \frac{d}{dt} \!+\! H 
\!+\! \frac{\dot{\Phi}}{1 \!+\! 2 \Phi}\Biggr] \Biggl[ \frac{ \overline{\tau} \!+\! 
\delta \tau_{00}}{1 \!+\! 2 \Psi} \Biggr] . \qquad
\end{eqnarray}
The fact that this simple result agrees with the direct computation is an
excellent check on accuracy.

\centerline{\bf Acknowledgements}

We are grateful for correspondence and conversations with C. Deffayet and S. Park. 
This work was partially supported by NSF grant PHY-1506513 and by the Institute for 
Fundamental Theory at the University of Florida.


\begin{thebibliography}{99}

\bibitem{Aprile:2017iyp} 
  E.~Aprile {\it et al.} [XENON Collaboration],
  Phys.\ Rev.\ Lett.\  {\bf 119}, no. 18, 181301 (2017)
  doi:10.1103/PhysRevLett.119.181301
  [arXiv:1705.06655 [astro-ph.CO]].

\bibitem{Cui:2017nnn} 
  X.~Cui {\it et al.} [PandaX-II Collaboration],
  Phys.\ Rev.\ Lett.\  {\bf 119}, no. 18, 181302 (2017)
  doi:10.1103/PhysRevLett.119.181302
  [arXiv:1708.06917 [astro-ph.CO]].

\bibitem{McGaugh:2016leg} 
  S.~McGaugh, F.~Lelli and J.~Schombert,
  Phys.\ Rev.\ Lett.\  {\bf 117}, no. 20, 201101 (2016)
  doi:10.1103/PhysRevLett.117.201101
  [arXiv:1609.05917 [astro-ph.GA]].

\bibitem{Lelli:2017vgz} 
  F.~Lelli, S.~S.~McGaugh, J.~M.~Schombert and M.~S.~Pawlowski,
  Astrophys.\ J.\  {\bf 836}, no. 2, 152 (2017)
  doi:10.3847/1538-4357/836/2/152
  [arXiv:1610.08981 [astro-ph.GA]].

\bibitem{Milgrom:1983ca} 
  M.~Milgrom,
  Astrophys.\ J.\  {\bf 270}, 365 (1983).
  doi:10.1086/161130

\bibitem{Milgrom:1983pn} 
  M.~Milgrom,
  Astrophys.\ J.\  {\bf 270}, 371 (1983).
  doi:10.1086/161131
        
\bibitem{Milgrom:1983zz} 
  M.~Milgrom,
  Astrophys.\ J.\  {\bf 270}, 384 (1983).
  doi:10.1086/161132

\bibitem{Woodard:2006nt} 
  R.~P.~Woodard,
  Lect.\ Notes Phys.\  {\bf 720}, 403 (2007)
  doi:10.1007/978-3-540-71013-4\_14
  [astro-ph/0601672].

\bibitem{Bekenstein:2004ne} 
  J.~D.~Bekenstein,
  Phys.\ Rev.\ D {\bf 70}, 083509 (2004)
  Erratum: [Phys.\ Rev.\ D {\bf 71}, 069901 (2005)]
  doi:10.1103/PhysRevD.70.083509, 10.1103/PhysRevD.71.069901
  [astro-ph/0403694].

\bibitem{Skordis:2005xk} 
  C.~Skordis, D.~F.~Mota, P.~G.~Ferreira and C.~Boehm,
  Phys.\ Rev.\ Lett.\  {\bf 96}, 011301 (2006)
  doi:10.1103/PhysRevLett.96.011301
  [astro-ph/0505519].

\bibitem{Skordis:2005eu} 
  C.~Skordis,
  Phys.\ Rev.\ D {\bf 74}, 103513 (2006)
  doi:10.1103/PhysRevD.74.103513
  [astro-ph/0511591].

\bibitem{Dodelson:2006zt} 
  S.~Dodelson and M.~Liguori,
  Phys.\ Rev.\ Lett.\  {\bf 97}, 231301 (2006)
  doi:10.1103/PhysRevLett.97.231301
  [astro-ph/0608602].

\bibitem{Bourliot:2006ig} 
  F.~Bourliot, P.~G.~Ferreira, D.~F.~Mota and C.~Skordis,
  Phys.\ Rev.\ D {\bf 75}, 063508 (2007)
  doi:10.1103/PhysRevD.75.063508
  [astro-ph/0611255].

\bibitem{Zlosnik:2007bu} 
  T.~G.~Zlosnik, P.~G.~Ferreira and G.~D.~Starkman,
  Phys.\ Rev.\ D {\bf 77}, 084010 (2008)
  doi:10.1103/PhysRevD.77.084010
  [arXiv:0711.0520 [astro-ph]].

\bibitem{Contaldi:2008iw} 
  C.~R.~Contaldi, T.~Wiseman and B.~Withers,
  Phys.\ Rev.\ D {\bf 78}, 044034 (2008)
  doi:10.1103/PhysRevD.78.044034
  [arXiv:0802.1215 [gr-qc]].
      
\bibitem{Freire:2012mg} 
  P.~C.~C.~Freire {\it et al.},
  Mon.\ Not.\ Roy.\ Astron.\ Soc.\  {\bf 423}, 3328 (2012)
  doi:10.1111/j.1365-2966.2012.21253.x
  [arXiv:1205.1450 [astro-ph.GA]].

\bibitem{Reyes:2010tr} 
  R.~Reyes, R.~Mandelbaum, U.~Seljak, T.~Baldauf, J.~E.~Gunn, L.~Lombriser and R.~E.~Smith,
  Nature {\bf 464}, 256 (2010)
  doi:10.1038/nature08857
  [arXiv:1003.2185 [astro-ph.CO]].

\bibitem{Dodelson:2011qv} 
  S.~Dodelson,
  Int.\ J.\ Mod.\ Phys.\ D {\bf 20}, 2749 (2011)
  doi:10.1142/S0218271811020561
  [arXiv:1112.1320 [astro-ph.CO]].

\bibitem{Boran:2017rdn} 
  S.~Boran, S.~Desai, E.~O.~Kahya and R.~P.~Woodard,
  arXiv:1710.06168 [astro-ph.HE].

\bibitem{Desai:2008vj} 
  S.~Desai, E.~O.~Kahya and R.~P.~Woodard,
  Phys.\ Rev.\ D {\bf 77}, 124041 (2008)
  doi:10.1103/PhysRevD.77.124041
  [arXiv:0804.3804 [astro-ph]].

\bibitem{Zlosnik:2006zu} 
  T.~G.~Zlosnik, P.~G.~Ferreira and G.~D.~Starkman,
  Phys.\ Rev.\ D {\bf 75}, 044017 (2007)
  doi:10.1103/PhysRevD.75.044017
  [astro-ph/0607411].

\bibitem{Sanders:2011wa} 
  R.~H.~Sanders,
  Phys.\ Rev.\ D {\bf 84}, 084024 (2011)
  doi:10.1103/PhysRevD.84.084024
  [arXiv:1105.3910 [gr-qc]].

\bibitem{Blanchet:2011wv} 
  L.~Blanchet and S.~Marsat,
  Phys.\ Rev.\ D {\bf 84}, 044056 (2011)
  doi:10.1103/PhysRevD.84.044056
  [arXiv:1107.5264 [gr-qc]].

\bibitem{Milgrom:2009gv} 
  M.~Milgrom,
  Phys.\ Rev.\ D {\bf 80}, 123536 (2009)
  doi:10.1103/PhysRevD.80.123536
  [arXiv:0912.0790 [gr-qc]].

\bibitem{Milgrom:2013iea} 
  M.~Milgrom,
  Phys.\ Rev.\ D {\bf 89}, no. 2, 024027 (2014)
  doi:10.1103/PhysRevD.89.024027
  [arXiv:1308.5388 [gr-qc]].
  
\bibitem{Woodard:2014wia} 
  R.~P.~Woodard,
  Can.\ J.\ Phys.\  {\bf 93}, no. 2, 242 (2015)
  doi:10.1139/cjp-2014-0156
  [arXiv:1403.6763 [astro-ph.CO]].
  
\bibitem{Wang:2015eaa} 
  C.~L.~Wang and R.~P.~Woodard,
  Phys.\ Rev.\ D {\bf 92}, 084008 (2015)
  doi:10.1103/PhysRevD.92.084008
  [arXiv:1508.01564 [gr-qc]].
  
\bibitem{Park:2015kua} 
  S.~Park, T.~Prokopec and R.~P.~Woodard,
  JHEP {\bf 1601}, 074 (2016)
  doi:10.1007/JHEP01(2016)074
  [arXiv:1510.03352 [gr-qc]].

\bibitem{Frob:2016fcr} 
  M.~B.~Fröb and E.~Verdaguer,
  JCAP {\bf 1603}, no. 03, 015 (2016)
  doi:10.1088/1475-7516/2016/03/015
  [arXiv:1601.03561 [hep-th]].

\bibitem{Deffayet:2011sk} 
  C.~Deffayet, G.~Esposito-Farese and R.~P.~Woodard,
  Phys.\ Rev.\ D {\bf 84}, 124054 (2011)
  doi:10.1103/PhysRevD.84.124054
  [arXiv:1106.4984 [gr-qc]].

\bibitem{Deffayet:2014lba} 
  C.~Deffayet, G.~Esposito-Farese and R.~P.~Woodard,
  Phys.\ Rev.\ D {\bf 90}, no. 6, 064038 (2014)
  Addendum: [Phys.\ Rev.\ D {\bf 90}, no. 8, 089901 (2014)]
  doi:10.1103/PhysRevD.90.089901, 10.1103/PhysRevD.90.064038
  [arXiv:1405.0393 [astro-ph.CO]].

\bibitem{Kim:2016nnd} 
  M.~Kim, M.~H.~Rahat, M.~Sayeb, L.~Tan, R.~P.~Woodard and B.~Xu,
  Phys.\ Rev.\ D {\bf 94}, no. 10, 104009 (2016)
  doi:10.1103/PhysRevD.94.104009
  [arXiv:1608.07858 [gr-qc]].

\bibitem{Ade:2015xua} 
  P.~A.~R.~Ade {\it et al.} [Planck Collaboration],
  Astron.\ Astrophys.\  {\bf 594}, A13 (2016)
  doi:10.1051/0004-6361/201525830
  [arXiv:1502.01589 [astro-ph.CO]].

\bibitem{Riess:2016jrr} 
  A.~G.~Riess {\it et al.},
  Astrophys.\ J.\  {\bf 826}, no. 1, 56 (2016)
  doi:10.3847/0004-637X/826/1/56
  [arXiv:1604.01424 [astro-ph.CO]].

\bibitem{Nojiri:2007uq} 
  S.~Nojiri and S.~D.~Odintsov,
  Phys.\ Lett.\ B {\bf 659}, 821 (2008)
  doi:10.1016/j.physletb.2007.12.001
  [arXiv:0708.0924 [hep-th]].

\bibitem{Deser:2013uya} 
  S.~Deser and R.~P.~Woodard,
  JCAP {\bf 1311}, 036 (2013)
  doi:10.1088/1475-7516/2013/11/036
  [arXiv:1307.6639 [astro-ph.CO]].

\bibitem{Woodard:2014iga} 
  R.~P.~Woodard,
  Found.\ Phys.\  {\bf 44}, 213 (2014)
  doi:10.1007/s10701-014-9780-6
  [arXiv:1401.0254 [astro-ph.CO]].

\bibitem{Soussa:2003vv} 
  M.~E.~Soussa and R.~P.~Woodard,
  Class.\ Quant.\ Grav.\  {\bf 20}, 2737 (2003)
  doi:10.1088/0264-9381/20/13/321
  [astro-ph/0302030].

\bibitem{Deser:2007jk} 
  S.~Deser and R.~P.~Woodard,
  Phys.\ Rev.\ Lett.\  {\bf 99}, 111301 (2007)
  doi:10.1103/PhysRevLett.99.111301
  [arXiv:0706.2151 [astro-ph]].

\bibitem{Arraut:2013qra} 
  I.~Arraut,
  Int.\ J.\ Mod.\ Phys.\ D {\bf 23}, 1450008 (2014)
  doi:10.1142/S0218271814500084
  [arXiv:1310.0675 [gr-qc]].

\bibitem{Deffayet:2009ca} 
  C.~Deffayet and R.~P.~Woodard,
  JCAP {\bf 0908}, 023 (2009)
  doi:10.1088/1475-7516/2009/08/023
  [arXiv:0904.0961 [gr-qc]].

\bibitem{Park:2012cp} 
  S.~Park and S.~Dodelson,
  Phys.\ Rev.\ D {\bf 87}, no. 2, 024003 (2013)
  doi:10.1103/PhysRevD.87.024003
  [arXiv:1209.0836 [astro-ph.CO]].

\bibitem{Dodelson:2013sma} 
  S.~Dodelson and S.~Park,
  Phys.\ Rev.\ D {\bf 90}, 043535 (2014)
  doi:10.1103/PhysRevD.90.043535
  [arXiv:1310.4329 [astro-ph.CO]].
                              
\bibitem{Nersisyan:2017mgj} 
  H.~Nersisyan, A.~F.~Cid and L.~Amendola,
  JCAP {\bf 1704}, no. 04, 046 (2017)
  doi:10.1088/1475-7516/2017/04/046
  [arXiv:1701.00434 [astro-ph.CO]].
  
\bibitem{Park:2017zls} 
  S.~Park,
  arXiv:1711.08759 [gr-qc].
  
\bibitem{Bekenstein:2008pc} 
  J.~D.~Bekenstein and E.~Sagi,
  Phys.\ Rev.\ D {\bf 77}, 103512 (2008)
  doi:10.1103/PhysRevD.77.103512
  [arXiv:0802.1526 [astro-ph]].
  
\bibitem{Milgrom:2008cs} 
  M.~Milgrom,
  Astrophys.\ J.\  {\bf 698}, 1630 (2009)
  doi:10.1088/0004-637X/698/2/1630
  [arXiv:0810.4065 [astro-ph]].

\end{thebibliography}
\end{document}